\documentclass[12pt,draftclsnofoot,onecolumn]{IEEEtran}
\IEEEoverridecommandlockouts
\usepackage{caption}
\usepackage{graphicx}
\usepackage{subfigure}
\usepackage{lipsum}
\usepackage{url}
\usepackage{verbatim}
\usepackage{todonotes}
\usepackage{color}

\usepackage{graphics}
\usepackage{amsmath,amsfonts, amssymb}
\usepackage{mathtools}
\usepackage{epstopdf}
\usepackage{amsthm}
\usepackage{bbm, bm}
\usepackage{algorithm}
\usepackage[noend]{algpseudocode}
\usepackage{dsfont}
\usepackage{cite}
\usepackage{multicol}
\usepackage{multirow}
\usepackage{array}
\usepackage{booktabs}
\usepackage[normalem]{ulem}
\useunder{\uline}{\ul}{}

\newtheorem*{remark}{Remark}

\newcolumntype{P}[1]{>{\centering\arraybackslash}p{#1}}
\newcolumntype{M}[1]{>{\centering\arraybackslash}m{#1}}

\ifCLASSINFOpdf
\else
\fi

\begin{document}
\title{All You Need is Feedback: Communication with Block Attention Feedback Codes}

\author{Emre~Ozfatura,
        Yulin~Shao,
        Alberto~Perotti,
        Branislav~Popovic,
        Deniz~G{\"u}nd{\"u}z
\thanks{E. Ozfatura, Y. Shao and D. G{\"u}nd{\"u}z are with Information Processing and Communications Lab, Department of Electrical and Electronic Engineering,
Imperial College London. Emails: \{m.ozfatura, y.shao, d.gunduz\} @imperial.ac.uk.}
\thanks{A. Perotti and B. Popovic are with the Radio Transmission Technology
Lab, Huawei Technologies Sweden AB, Kista 164-94, Sweden. Emails:
\{alberto.perotti, branislav.popovic\}@huawei.com}}

\maketitle

\begin{abstract}
Deep neural network (DNN)-based channel code designs have recently gained interest as an alternative to conventional coding schemes, particularly for channels where existing codes do not provide satisfactory performance. Coding in the presence of feedback is one such problem, for which promising results have recently been obtained by various DNN-based coding architectures. In this paper, we introduce a novel learning-aided feedback code design, dubbed \textit{generalized block attention feedback (GBAF) codes}, that achieves order-of-magnitude improvements in block error rate (BLER) compared to existing solutions. Sequence-to-sequence encoding and block-by-block processing of the message bits in GBAF codes not only reduce the communication overhead due to reduced number of interactions between the transmitter and receiver, but also enable flexible coding rates. More importantly, GBAF codes provide a modular structure that can be implemented using different neural network architectures. In this work, we employ the transformer architecture, which outperforms all the prior DNN-based code designs in terms the block error rate in the low signal-to-noise ratio regime when the feedback channel is noiseless. 
\end{abstract}


% \fi
\IEEEpeerreviewmaketitle

\section{Introduction}
Reliable communication in the presence of noise has been a long-standing challenge. Numerous coding and modulation techniques have been invented over many decades to push the boundaries of communication; that is, to achieve higher data rates with less error probability under given resource constraints (bandwidth, power).
Information storage and communication are two core technologies that underpin the information age, and the success of both hinges on error correction codes, such as BCH, Reed-Muller, convolution, turbo, low-density parity-check (LDPC), and polar codes. While these codes can approach the fundamental Shannon capacity limit over an additive white Gaussian noise (AWGN) channel in the large blocklength regime, there are many scenarios where we do not have practical codes that approach the fundamental theoretical boundaries. 

Coding in the presence of feedback is one such challenging, yet practical scenario. The classical feedback channel model was introduced and studied by Shannon \cite{Shannon}. In general, the formulation of communication with feedback involves a transmitter-receiver pair connected via a forward and a feedback channel, and the goal is to reliably deliver a block of bits from the transmitter to the receiver with the help of feedback. Shannon investigated the impact of feedback on the forward channel capacity by assuming perfect channel output feedback with unit delay. He proved an important result that the classical capacity of a memoryless forward channel does not increase in the presence of feedback \cite{Shannon}. 

While feedback does not increase the capacity, it is known to simplify the communication scheme and improve the reliability in the finite blocklength regime. For example, most practical communication systems involve feedback either in the form of channel state information feedback, or automatic repeat requests (ARQs). While the former simply provides adaptation to channel variations, the latter increases reliability by adjusting the codelength according to the noise realization. Another method to exploit feedback to increase reliability was introduced by Schalkwijk and Kailath in \cite{SK1} and \cite{SK2}. In the classical Schalkwijk-Kailath (SK) scheme, the transmitter encodes its message using pulse amplitude modulation (PAM) initially, and subsequently refines the estimate of the message at the receiver in an iterative manner by sending a scaled version of the residual error at each iteration. Provided that the transmission rate is below the capacity, the SK scheme achieves a double exponential decay of the decoding error probability with the increase in code length. Designing coding and modulation schemes that can best exploit the feedback has been an ongoing challenge over decades \cite{Shannon, SK1, SK2,UBfeedback, Weissman2008, Gallager, Love2011, Kim2007ISIT,  ModuloSK, deepcode,defc, drf, AttentionCode}, yielding a significant impact on a variety of applications that require ultra-reliable short-packet communications \cite{URLLC}, such as autonomous vehicles, industrial automation and control, tactile Internet, and augmented/virtual reality, to count a few.

Existing feedback codes can be classified as `human-crafted' codes \cite{SK1, SK2, Weissman2008, Love2011, Kim2007ISIT, ModuloSK}, and deep learning (DL)-aided codes \cite{deepcode, defc, drf, AttentionCode}. Among human-crafted codes, two notable works are the SK scheme \cite{SK1,SK2,Gallager} and its extension to the \textit{active feedback} scenario, the modulo-SK scheme \cite{ModuloSK}. Here, active feedback refers to the scenario in which the feedback symbols can also be encoded by the receiving terminal prior to transmission to the transmitter.

%The essence of these two schemes is describing the block of bits by a $2^K$-ary pulse-amplitude modulation (PAM) constellation, where $K$ is the blocklength, and transmitting the constellation point corresponding to the message iteratively over many interactions, where each interaction consists of one forward and one feedback transmission. At each interaction, the transmitter first recovers the receiver's estimate of the constellation point, thanks to the feedback signal, and transmits a scaled signal to correct the estimation error in the next interaction. 

% the receiver feeds back its estimate of the constellation to the transmitter such that the transmitter can compensate for the receiver's estimation error in the next interaction.
A main disadvantage of the aforementioned human-crafted codes is that they are sensitive to numerical precision and quantization errors \cite{Gallager, Kim2007ISIT, deepcode, AttentionCode}. Since the message is mapped to a $2^{K}$-ary PAM constellation, the number of bits required to represent all the statistics in this process grows linearly with $K$. When $K$ is large, these schemes suffer from severe quantization errors caused by the finite-precision arithmetic and finite quantization levels of the electronic parts and components, e.g., power amplifier and FPGA chip. On the other hand, DL-aided feedback codes model the communication system as an autoencoder \cite{deepcode,defc,drf,AttentionCode}, in which the encoder and decoder are modeled as a pair of deep neural networks (DNNs), while the wireless channel is treated as an untrainable stochastic layer. The code is obtained by end-to-end unsupervised learning to minimize the reconstruction error of the block of bits at the receiver.

Compared with human-crafted codes, DL-based feedback codes do not suffer from the constraint of finite precision and quantization levels, as they can be trained with such constraints embedded into the training process. Moreover, they are very flexible and can be easily trained for different scenarios. Specifically, both the SK and modulo-SK schemes are designed for the setup of unit-time delayed feedback and AWGN channels with a specific pair of feedforward and feedback signal-to-noise ratios (SNRs). In contrast, DL-based codes can be easily generalized to more practical scenarios \cite{deepcode,AttentionCode}, such as feedback with greater delays, block feedback, as well as non-Gaussian noise or fading channels.
On the other hand, existing DL-aided feedback codes  suffer from the following limitations that we address in this paper: 
\begin{itemize}
\item \textit{Communication overhead:} In practice, each round of feedback subsequent to the use of the forward channel introduces an overhead and additional delay independent of the number of transmitted bits. We quantify the corresponding communication overhead as the number of ``switches'' at the source node, between transmitting parity symbols and receiving feedback symbols, or equivalently the number of communication rounds $T$. In the previous designs, $T$ scales linearly with with the number of message bits $K$. One of our key objectives is to reduce this communication overhead without sacrificing performance significantly.

\item \textit{Limited set of feasible rates:} Existing schemes are limited to code rates of $1/k, k\in\mathbb{Z}^{+}$. Hence, another important aspect of this work is to present a design that can transmit at a wider range of rates. The flexibility in the communication rate is important to achieve higher spectral efficiencies, particularly in the higher SNR regimes.

\item \textit{Lack of structure:} Existing codes are defined through the employed DNN architecture. Instead, we would like to provide a holistic view of the problem and introduce a generalized modular design, where modules can be added/removed, and implemented through arbitrary architectures addressing different requirements in terms of performance and complexity. 
\end{itemize}

In this paper, we introduce the generalized block attention feedback (GBAF) code, which addresses all of the aforementioned limitations of existing designs. In particular, in the GBAF architecture, we introduce a novel sequence-to-sequence encoding framework. We then group the message bits into blocks, and treat each block as the information unit to be communicated. We employ the popular transformer-based encoder architecture \cite{attention1, attention2, Wang:transformer:comms} as its core sequence-to-sequence encoder module, GBAF codes achieve orders of magnitude improvements in terms of the BLER performance over the whole range of channel SNRs compared to existing DL-based codes in the literature. Apart from \cite{ModuloSK}, feedback codes in the literature are designed for a passive feedback scenario; that is, the feedback signal is simply a noisy version of the signal received at the receiver. While we also consider passive feedback in this paper, our design can be easily extended to active feedback.

The rest of the paper is organized as follows. We present the problem formulation in Section \ref{s:problem}, and provide a detailed overview of the existing feedback code structures and there limitations. The structure and modules of the GBAF code are introduced in Section \ref{s:GBAFcodes}. Numerical results illustrating its superiority are presented in Section \ref{s:experiments}. We conclude the paper in Section \ref{s:Conclusion}.

%The introduced design allows different variations from the operational point of view such as using systematic or unsystematic coding, adding/removing blocks, enabling/disabling feedback connections as well being able to be used with any sequence-to-sequence encoding NN architecture.

{\it Notations} -- We use bold, capital bold, and capital calligraphic fonts to denote vectors, matrices, and sets, respectively, i.e., $\bm{v}$, $\bm{V}$, and $\mathcal{V}$. We use the notation $\bm{v}_{[~]}$, $\bm{V}_{[~,~]}$ to denote index slicing. We use the superscript for a vector/matrix/list to refer to its realization at a particular time/iteration.  Finally, we use subscripts to emphasize a particular element of a sequence; for example, given a sequence of vectors $\mathcal{Q}=\left\{\bm{q}_{1},\ldots,\bm{q}_{K}\right\}$, $\bm{q}_{i}$ is used to represent the $i$th vector in the sequence.

\section{Problem Statement}\label{s:problem}
\subsection{System model}
\begin{figure}[t]
  \centering
  \includegraphics[width=0.5\columnwidth]{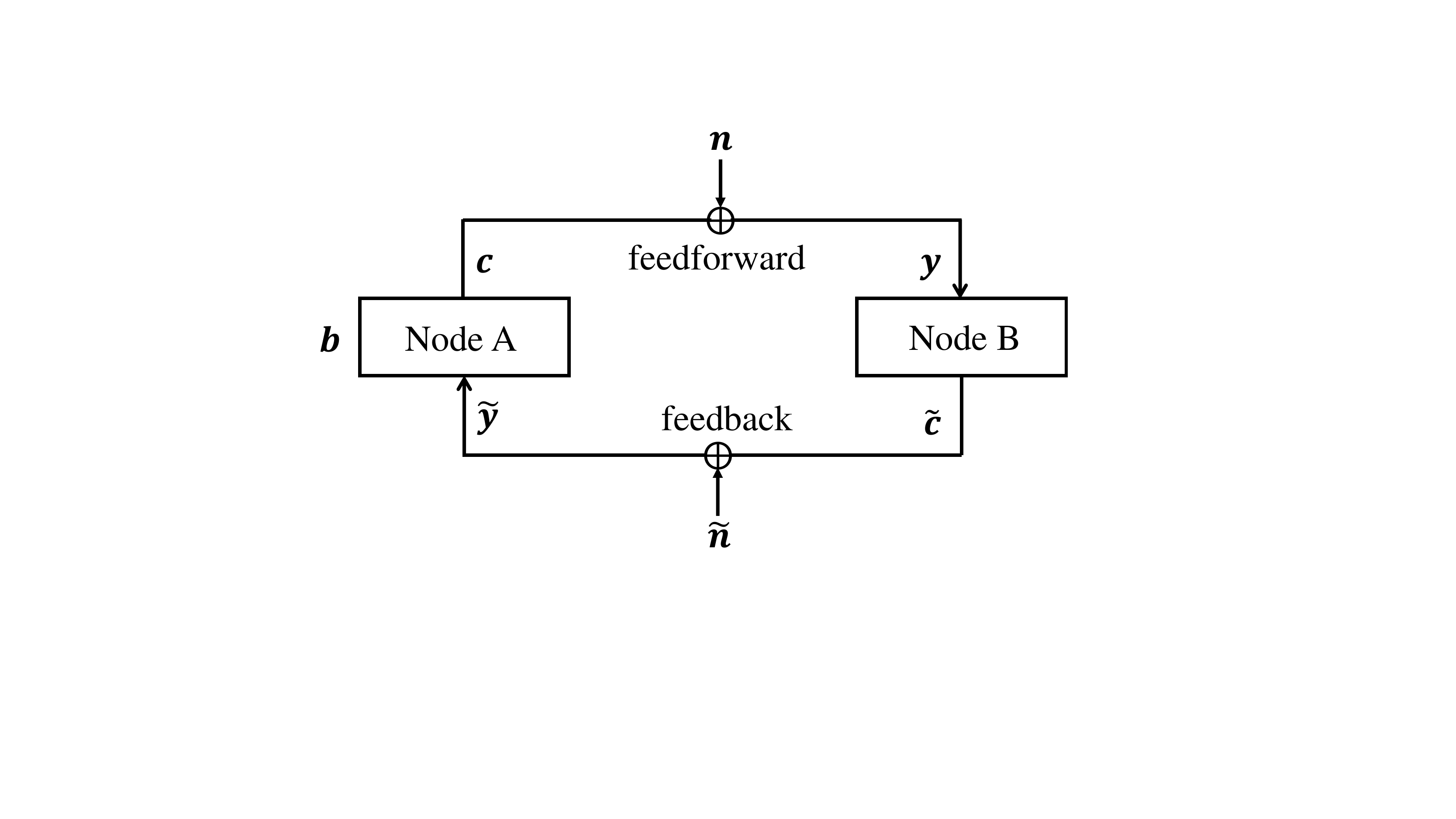}\\
  \caption{Communication with block feedback: the system model. }
\label{fig:model}
\end{figure}

We consider a point-to-point communication scenario with one transmitter and one receiver, as shown in Fig.~\ref{fig:model}. The objective of the transmitter is to send $K$ bits of information, $\bm{b}=[b_{1},\ldots,b_{K}]\in \left\{0,1\right\}^{K}$, to the receiver in $N$ channel uses. We impose a rate constraint of $R$, that is, $K/N \geq R$. Here, we use $\bm{c}=[c_{1},\ldots,c_N]\in \mathbb{R}^N$ to denote the sequence of transmitted symbols over the forward channel. We model both the forward and feedback channels as AWGN channels with independent noise terms. 

%In particular, the symbol at the receiver at the $t$-th channel use,  $y_{t}$, is given by
%\begin{equation}\label{noise}
%y_{t} = c_{t} + n_{t},
%\end{equation}
%where $n_{t}\sim(0,\sigma_{f}^{2})$ are independent Gaussian noise samples. Accordingly, we use $\bm{y}=[y_{1},\ldots,y_{N}]$ to denote the sequence of received symbols at the end of $N$ channel uses. %Up to now we have describe  communication at forward direction; next we explain the key aspects of communication with feedback.
%\subsection{Communication with feedback}

%To provide a better understanding of the feedback mechanism, as well as to understand how different variations can be obtained, we introduce the term {\em communication block}, which describe the interval of consecutive channel uses, where only one node is actively transmitting and only one node is actively receiving symbols. 

We consider feedback model consisting of multiple communication rounds, where in each round the transmitter transmits a vector of symbols, after which it receives a vector of feedback symbols corresponding to the transmitted symbols over the forward channel. This is in contrast to the commonly considered model, where the transmitter receives a feedback symbol corresponding to each transmitted symbol with unit delay. Our model would be particularly relevant in the active feedback scenario, where the feedback symbols are encoded by the receiver. In the case of passive feedback considered in this paper, we use this model to quantify the potential overheads due to processing of the feedback symbols and generating the transmitted symbols over the forward channel based on the received feedback. In the literature, the channel output feedback, noiseless or noisy, is assumed to be available instantly at the encoder. However, in practice, these feedback symbols need to be encoded and/or modulated as well, and in general, encoding/ decoding operations, as well as the additional exchange of control information between the transmitter and receiver for every forward and feedback packet will introduce additional overheads. Hence, in practice, it is desired to utilize feedback while introducing minimum overhead. Therefore, our goal will be to achieve the desired level of reliability with minimal number of interactions, i.e., communication rounds, between the transmitter and the receiver.

%we would like to transmit only a few packets in the forward and feedback channels, although this would reduce the amount of feedback resourced exploited (since no feedback is received for the last transmitted block). 

Let $\tau$ denote the index of the \textit{communication round}. In communication round $\tau$, the transmitter sends $N_{\tau}$ symbols, denoted by $\bm{c}^{(\tau)}$, in the forward direction, and receives $N_{\tau}$ symbols\footnote{In general, in the active feedback scenario, we can have a different number of symbols transmitted over each communication round of the forward and feedback channels. Here, we set them to be equal as we assume that the symbols transmitted over the feedback channel are simply the symbols received by the receiver, i.e., passive feedback.}, denoted by $\tilde{\bm{y}}^{(\tau)}$, over the feedback link, for $\tau=1,\ldots,T-1$. We have $\sum_{\tau=1}^{T}N_{\tau}\leq N$. We remark that often the existing schemes, as well as the proposed design, utilize equal length vectors over $\tau$, where a slight modification appears in the systematic code design used in the previous works, which we will explain later. The communication is terminated when the receiver receives $\bm{c}^{(T)}$. %Due to the noise additive nature of the channel, as described in (\ref{noise}), 
The received vector of symbols at the forward and feedback links, denoted by $\bm{y}^{(\tau)}$ and $\tilde{\bm{y}}^{(\tau)}$, respectively, are given by
\begin{equation}
\bm{y}^{(\tau)} = \bm{c}^{(\tau)} + \bm{n}^{(\tau)},~~~\mbox{for } \tau=1, \ldots, T,
\end{equation}
and
\begin{equation}
\tilde{\bm{y}}^{(\tau)} = \bm{y}^{(\tau)} + \tilde{\bm{n}}^{(\tau)},~~~\mbox{for } \tau=1, \ldots, T-1,
\end{equation}
where $\bm{n}^{(\tau)}, \tilde{\bm{n}}^{(\tau)} \in \mathbb{R}^{N_{\tau}}$ are the noise vectors consisting of independent and identically distributed (i.i.d.) zero-mean Gaussian random variables with variances $\sigma^{2}_{ff}$ and $\sigma^{2}_{fb}$, respectively.

% Similarly, a constraint on the feedback direction can be introduced as,
% \begin{equation}
% \sum_{\tau=1}^{T-1}\tilde{N}_{\tau}\leq \tilde{N}.
% \end{equation}

If we consider $T$ communication rounds in the forward direction, this also implies that the direction of communication is switched $T$ times, which corresponds to the overhead of the feedback mechanism. As mentioned above, larger $T$ corresponds to more overhead. 

%For example, if $T=1$, there is only one forward transmission of $N$ symbols. On the other hand, if $N_{\tau}=1$, $\forall \tau$, then we have $N$ blocks and the transmitter waits for a feedback transmission after each forward transmission.

% In the general form of the described protocol, $N_{\tau}$ and  $\tilde{N}_{\tau}$ can be chosen as desired. For example, one may want to keep $\tilde{N}_{\tau}$ small to avoid potential delay due to feedback. Nevertheless, we consider $N_{\tau}=\tilde{N}_{\tau}$ in the scope of this work. This choice is often desired in order to align forward and feedback symbols for the encoding mechanism.\\

The focus of our paper is to design a mechanism for generating symbols in forward and feedback directions for each  communication round $\tau$. Before describing the particular encoding mechanism we propose, we introduce the so-called `\textit{knowledge vectors}' $\boldsymbol{q}^{(\tau)}$ and $\tilde{\boldsymbol{q}}^{(\tau)}$, which refer to all the available information at the transmitter and the receiver, respectively, when generating the symbols transmitted in communication round $\tau$. The knowledge vector at the transmitter, $\boldsymbol{q}^{(\tau)}$, consists of the original bit stream, previously transmitted symbols, and the received feedback symbols up to time $\tau$, i.e.,
\begin{equation}
\boldsymbol{q}^{(\tau)} =[\bm{b},\bm{c}^{(1)},\ldots,\bm{c}^{(\tau-1)}, \tilde{\bm{y}}^{(1)},\ldots,\tilde{\bm{y}}^{(\tau-1)}].
\end{equation}
The knowledge vector at the receiver consists of the received channel outputs up to time $\tau$ 
\begin{equation}
\tilde{\boldsymbol{q}}^{(\tau)} =[ \bm{y}^{(1)}, \ldots, \bm{y}^{(\tau)}].
\end{equation}
% \begin{equation}
% \tilde{\boldsymbol{q}}^{(\tau)} =[\tilde{\bm{c}}^{(1)},\ldots,\tilde{\bm{c}}^{(\tau-1)}, \bm{y}^{(1)},\ldots,\bm{y}^{(\tau)}].
% \end{equation}

% Let  $M^{({\tau})}$ and $\tilde{M}^{({\tau})}$ denote the encoding mechanisms for the forward and feedback directions, respectively, such that at each communication round $\tau$ we have the following mappings
% \begin{equation}
% M^{({\tau})}:  \boldsymbol{q}^{(\tau)} \xrightarrow{} \bm{c}^{(\tau)}\in \mathbb{R}^{N_{\tau}},
% \end{equation}
% and 
% \begin{equation}
% \tilde{M}^{({\tau})}:  \tilde{\boldsymbol{q}}^{(\tau)} \xrightarrow{} \tilde{\bm{c}}^{(\tau)}\in \mathbb{R}^{\tilde{N}_{\tau}}.
% \end{equation}

{\color{red} 
\begin{table}[!h]
\caption{Notations}
\label{tab:notations}
\centering
\begin{tabular}{|l|l|}
\hline
\textbf{Notation} & \textbf{Description} \\ \hline
$\bm{b}$ & Input bit-stream \\ \hline
$K$ & Length of the bit-stream \\ \hline
$N$ & Codeword length \\ \hline
$R$ & Transmission rate \\ \hline
$T$ & Number of interactions (communication rounds) \\ \hline
$\tau$  & Index of communication rounds  \\ \hline
$\bm{c}^{(\tau)}$ & Coded symbols in forward direction in communication round $\tau$ \\ \hline
$\bm{y}^{(\tau)}$ & Received channel output at receiver in communication round $\tau$ \\ \hline
%$\tilde{\bm{c}}^{(\tau)}$ & Feedback symbol in round $\tau$ \\ \hline
$\tilde{\bm{y}}^{(\tau)}$ & Received channel feedback at transmitter in communication round $\tau$  \\ \hline
$\bm{q}^{(\tau)}$ & Knowledge vector at transmitter in communication round $\tau$  \\ \hline
$\bm{\tilde{q}}^{(\tau)}$ & Knowledge vector at receiver in communication round $\tau$  \\ \hline
\end{tabular}
\end{table}
}

Let  $M^{({\tau})}$ denote the encoding function at the transmitter, where $M^{({\tau})}(\boldsymbol{q}^{(\tau)}) = \bm{c}^{(\tau)}\in \mathbb{R}^{N_{\tau}}$. Once the transmission of all the symbols is completed, a decoding function $D$ is employed at the receiver to recover the original bit stream, i.e., $\bm{\hat{b}}\in\left\{0,1\right\}^{K} = D(\tilde{\boldsymbol{q}}^{(T)})$.

%\subsection{Power constraint}
The code must satisfy an average power constraint on the transmitted symbols:
\begin{align}
& \mathbb{E}\left[\frac{1}{N} \sum^{T}_{\tau=1}\langle \bm{c}^{(\tau)} , \bm{c}^{(\tau)} \rangle\right]\leq 1.
% & \mathbb{E}\left[\frac{1}{\tilde{N}} \sum^{T-1}_{\tau=1}\langle \tilde{\bm{c}}^{(\tau)} , \tilde{\bm{c}}^{(\tau)} \rangle\right]\leq 1\\
\end{align}
Hence, the SNR in the forward direction is given by $SNR_{ff}=1/\sigma_{ff}^{2}$, while the SNR in the feedback channel is $SNR_{fb}=1/\sigma_{fb}^{2}$. We refer to the case $\sigma_{fb}=0$ as {\em noiseless feedback}.

\begin{remark}[Systematic codes]
We refer to a feedback code as a {\em systematic feedback code}, if there is an additional initial stage at $\tau=0$, such that the encoder maps the original bit stream to its BPSK modulated version, i.e., $N_{0}=K$, and $M^{({0})}(\bm{b})= \bm{c}^{(0)} = \alpha(2 \cdot \bm{b}-1)$, where $\alpha$ is chosen to satisfy the power constraint.
\begin{equation}\label{encr}
M^{({0})}:  \boldsymbol{q}^{(0)}=\bm{b} \xrightarrow{BPSK} \bm{c}^{(0)} = \bar{\bm{b}}=2*\bm{b}-1.
\end{equation}
% and the encoder at the receiver simply relays the received noisy symbols, i.e.,
% $\tilde{N}_{1}=K$, and
% \begin{equation}\label{enct}
% \tilde{M}^{({1})}:  \tilde{\boldsymbol{q}}^{(1)}=\bm{y}^{(1)} \xrightarrow{relay} \tilde{\bm{c}}^{(1)} = \bar{\bm{b}}+\bm{n}^{(1)}.
% \end{equation}
We note that,  an additional iteration index $\tau=0$ is allocated for the systematic code part to be able to align different DL-based feedback code designs. Independent from the employment of the systematic part, DL-based symbol encoding starts from $\tau \geq 1$.

Note also that, although we have restricted the above definition to BPSK modulation for the sake of simplicity, the same notion can be extended to other modulation schemes with larger constellations. In general, there is no particular reason to restrict ourselves to a systematic feedback scheme, but we defined this set of codes explicitly as the DL-based codes considered in the literature \cite{deepcode, defc, drf} are all systematic codes.
\end{remark}

\subsection{Existing DL-Based Feedback Codes}
The ultimate challenge in feedback codes is designing an iterative encoding process for the parity symbols at the transmitter, and a decoding process for the received symbols at the receiver. DNN-based feedback codes aim to tackle this issue by considering the encoding and decoding mappings, $M^{(\tau)}$ and $D^{(\tau)}$, respectively, as DNN architectures, and by training them for a sufficient number of randomly generated bit streams to achieve the final network model/weights. It has been shown that such an end-to-end training approach is highly effective for designing feedback codes \cite{deepcode, drf, AttentionCode}. Now we revisit some of the existing feedback code designs in the literature and illustrate how they operate according to our generic framework.\\

\begin{figure}[h]
\begin{center}
\includegraphics[scale =0.45]{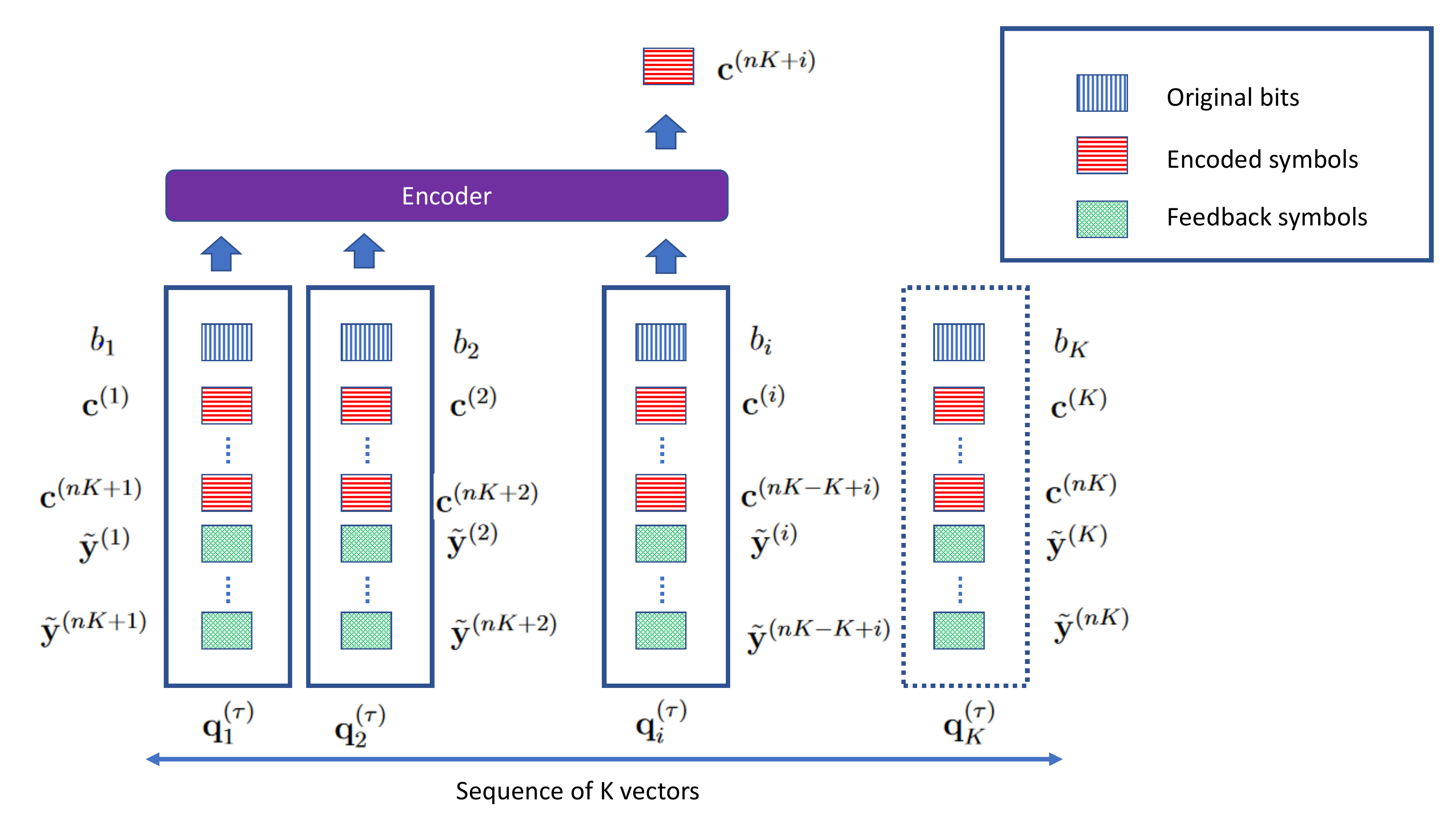}
\caption{Visualisation of the sequence-to-one encoding approach at iteration $\tau=nK+i$. The knowledge vector $\bm{q}^{(\tau)}$ is divided into $K$ parts, where $\bm{q}_{i}^{(\tau)}$ corresponds to the knowledge vector about the $i$-th message bit. Each generated channel input vector $\bm{c}^{(\tau)}$, where $\tau = nK+i$ corresponds to a particular message bit $i$, and only the knowledge vectors corresponding to message bits $1, \ldots, i$ are used to generate this channel input vector. The transmitted symbol and the corresponding channel output feedback are then added to the knowledge vector of bit $i$, $\bm{q}_{i}^{(\tau)}$, to be used in the generation of future channel symbols.}
\label{seq2one}
\end{center}
\end{figure}

\subsubsection{General overview}
All the existing DL-based feedback codes in the literature, the DeepCode \cite{deepcode}, the DEF code \cite{defc}, and the DRF code \cite{drf}, consider systematic and passive feedback schemes. The communication process is divided into two phases, $\tau=0$ and  $\tau>0$.
$M^{(0)}$ corresponds to the systematic modulation scheme described in (\ref{encr}). In the second phase, $\tau>0$, a DNN architecture, denoted by $H_{\mathrm{encoder}}$, is used as the encoder to generate the vector of parity symbols, i.e., we have
\begin{equation}
H_{\mathrm{encoder}}: S_{\mathrm{encoder}}(\bm{q}^{(\tau)}) \xrightarrow{Neural-encoder} \bm{c}^{(\tau)},
\end{equation}
where $S_{\mathrm{encoder}}(\cdot)$ denotes the pre-processing function that defines how the knowledge vector $\bm{q}^{(\tau)}$ is fed to the DNN architecture $H_{\mathrm{encoder}}$.

\subsubsection{Sequence-to-one encoding}
Although the existing DL-based encoder designs employ different NN architectures, see Table \ref{designs}, they all follow the same structure for processing the knowledge vector $\bm{q}^{(\tau)}$ in order to generate channel symbols. %Before providing further details, we want to clarify certain notions that will be used throughout the paper, and associate certain DL architectural concepts with our problem. We will use the term {\em sequence} to refer to an input structure that consists of a number of ordered elements, for instance, in natural language processing (NLP) a sentence is a sequence of words, and each word is an element of this sequence. In the context of our problem, one can also consider a bit-stream as a sequence consisting of bits. The main motivation behind using the notion of sequence is that the existing DL-based encoder architectures are originally designed for processing sequences in the ML literature.\\
Function $S_{\mathrm{encoder}}(\cdot)$ is used to transform the knowledge vector into a sequence of vectors that can be fed to the DL-based encoder. Hence, for the encoding process, $\bm{q}^{(\tau)}$ is first transformed into a sequence of vectors $\left\{\bm{q}_{1}^{(\tau)}, \ldots, \bm{q}_{K}^{(\tau)} \right\}$, whose length is equal to the length of the original bit-stream, which is then fed to the network to generate a vector of channel symbols. 

The encoding strategy, followed in the previous code designs, simply assumes that  $\bm{q}_{i}^{(\tau)}$ is the knowledge vector at round $\tau$ corresponding to the $i$th bit of the original bit-stream. The existing code designs, using sequence-to-one encoding approach, has two distinguishing features. First, at any communication round $\tau=nK+i$, during the $n+1$th pass over the bit-stream, they generate one vector of symbols $\bm{c}^{(\tau)}$ that corresponds to a particular knowledge vector $\bm{q}_{i}^{(\tau)}$; hence, when the feedback is available at the transmitter, only $\bm{q}_{i}^{(\tau)}$ is updated to obtain $\bm{q}_{i}^{(\tau+1)}$ before the next vector of symbols,  $\bm{c}^{(\tau+1)}$, is generated. Second, the encoding process is causal, that is for generating $\bm{c}^{(\tau)}$, $\tau=nK+i$, only knowledge vectors  $\left\{\bm{q}_{1}^{(\tau)}, \ldots, \bm{q}_{i}^{(\tau)}\right\}$ are utilized, simply those ones whose index is larger then $i$ are ignored. We illustrate the overall  sequence-to-one encoding process for a particular $\tau\geq1$ in  Fig. \ref{seq2one}.

While we provided above the general structure of the existing DL-based feedback code designs, they all use a special case of this general form with a single pass over the bit-stream, i.e., $n=1$,  and exactly two symbols are generated at each iteration, i.e., $N_{\tau}=2$ for all $\tau \geq1$ while systematic encoding with BPSK modulation is used for $\tau =0$. For this particular setup, one achieves the rate $R=1/3$ by using $K+1$ communication rounds in the forward direction. In general, for given $K$, $R$, and $N_{\tau}$, $T= \frac{K}{N_{\tau}}*\left(\frac{1}{R}-1\right)+1$ communication rounds are required in the forward direction\footnote{To prevent any confusion, we ignore the extra zero padding strategy introduced in \cite{deepcode}.}.

Similarly, at the receiver a combination of DNN architecture $H_{\mathrm{decoder}}$ and pre-processing function $S_{\mathrm{decoder}}(\cdot)$ is used as the decoding function $D$, i.e.,
\begin{equation}
H_{\mathrm{decoder}}\left(S_{\mathrm{decoder}}\left(\tilde{\boldsymbol{q}}^{(T)}\right)\right) = H_{\mathrm{decoder}}\left(S_{\mathrm{decoder}}\left( \bm{y}^{(1)},\ldots,\bm{y}^{(T)} \right)\right) = \hat{\bm{b}}\in\left\{0,1\right\}^{K}.
\end{equation}
The particular DNN architectures employed, both at the encoder and the decoder, in the existing feedback codes proposed in the literature are listed in Table \ref{designs}.

\begin{table}
\begin{center}
\begin{tabular}{ c c c }
\textbf{Design} & $H_{\mathrm{decoder}}$ & $H_{\mathrm{encoder}}$\\
\hline
 DeepCode \cite{deepcode} & Bi-GRU & GRU \cite{gru} \\ 
 \hline
 DRF Code \cite{drf} & Bi-LSTM & LSTM \cite{lstm}\\
  \hline
 AttentionCode \cite{AttentionCode} & Transformer Encoder & Transformer Encoder \cite{attention1}\\ 
 \hline
\end{tabular}
\end{center}
\caption{DNN-based designs for feedback codes.}
\label{designs}
\end{table}

\section{Generalized Block Attention Feedback (GBAF) Codes}\label{s:GBAFcodes}

Following the general design principles summarized above, the common aspect of the existing DNN-based feedback codes is to consider the given bit-stream as a sequence and utilize DNN architectures that are particularly designed for processing sequences, such as the long short-term memory (LSTM) and gate recurrent unit (GRU) architectures, to generate parity symbols as well as to decode them to recover the original bit stream. The proposed GBAF code design differentiates itself from the existing codes in several aspects. Below we present the architecture of the GBAF codes, and emphasize its main novelties with respect to the state-of-the art. 

\subsection{Overview of Innovations}
\subsubsection{\textbf{Sequence-to-sequence encoding}}
 The key novelty of the GBAF code design, different from the existing strategies, is the way the sequence is processed at the encoder to generate channel symbols. To clarify, all the existing strategies follow the {\em sequence-to-one} coding principle mentioned above, whereas the GBAF code uses {\em sequence-to-sequence} encoding to generate a vector of symbols corresponding to the while input bit stream. Similarly to sequence-to-one coding, the knowledge vector at the encoder, $\bm{q}$, is first transformed into a sequence of vectors $\left\{\bm{q}_{1}^{(\tau)}, \ldots, \bm{q}_{K}^{(\tau)}\right\}$. However, unlike the sequence-to-one coding, at each communication round $\tau$, parallel processing of these knowledge vectors is used instead of casual processing; that is for each knowledge vector $\bm{q}_{i}^{(\tau)}$, $i=1, \ldots, K$, a vector of coded symbols $\bm{c}^{(\tau)}_{i}$ is generated simultaneously and the transmitted codeword $\bm{c}^{(\tau)}$ is obtained by concatenating these vectors, i.e., $\bm{c}^{(\tau)} = \left[\bm{c}^{(\tau)}_{1},\ldots,\bm{c}^{(\tau)}_{K}\right]$. Accordingly, when the feedback is available, unlike the previous approach, all the elements of the sequence of knowledge vectors are updated simultaneously. Hence, the number of interactions between the receiver and the transmitter does not scale with the length of the sequence, but with the coding overhead, which is the inverse of the rate.

\begin{figure}[t]
\begin{center}
\includegraphics[scale = 0.45]{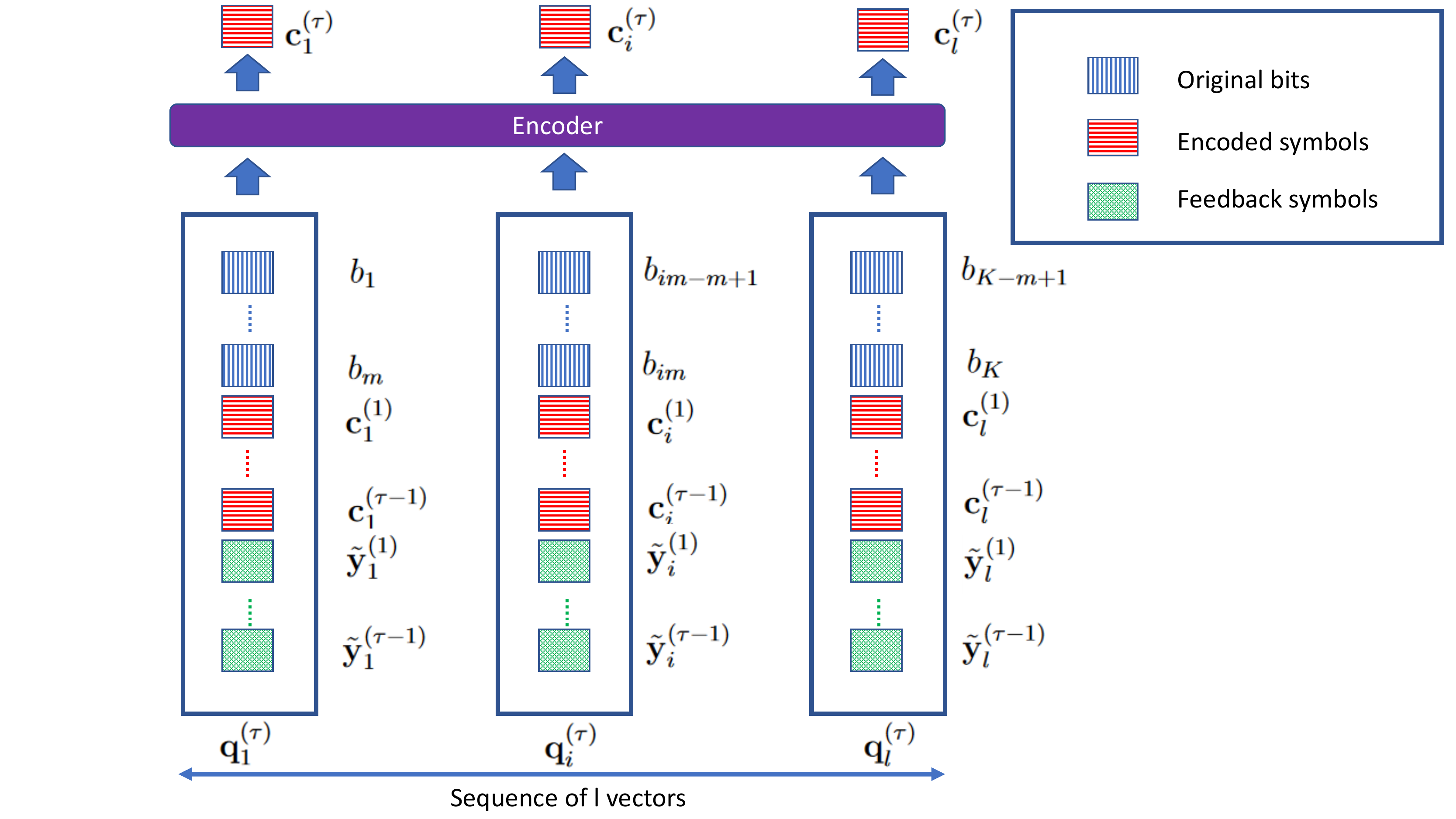}
\caption{ Visualisation of the sequence-to-sequence encoding approach with block formation for a block size of $m$, where $l=K/m$, at iteration $\tau$. Bits are grouped into $l$ blocks, each consisting of $m$ bits. The knowledge vector $\bm{q}^{(\tau)}$ is also divided into $l$, where $\bm{q}_{i}^{(\tau)}$ corresponds to the knowledge vector about the $l$-th block of message bits. Each generated channel symbol vector $\bm{c}_i^{(\tau)}$ corresponds to a particular block of message bits $b_{im-(m-1)}, \ldots, b_{im}$, but all the available knowledge vectors are used simultaneously to generate all the channel input vectors in iteration $\tau$. The transmitted symbol and the corresponding channel output feedback are then added to the knowledge vectors of all the message blocks.}
\label{seq2seq}
\end{center}
\end{figure}

\subsubsection{\textbf{Sequence of bits to sequence of blocks}}
Although the parallel execution with sequence-to-sequence encoding reduces the communication overhead compared to existing frameworks, the limitation of the number of feedback iterations by the rate leads to under-utilization of the feedback mechanism. Apart from that, we identified three other limitations:
\begin{enumerate}
    \item performing sequence-to-sequence encoding for large sequences is computationally expensive and has large memory requirements;
    \item  when the number of feedback iterations is limited, the information gathered for each element of the sequence is also limited; 
    \item the sequence-to-sequence encoding alone does not offer a wider range of rate options compared to former designs.
\end{enumerate}
 
 We want to highlight that, in the existing DL-based feedback codes, the length of the processed sequence is equal to the length of the bit-stream; that is, each element of the sequence corresponds a single message bit and the corresponding channel input and feedback symbols. Hence, to address all the aforementioned limitations, we divide the bit-stream into group of bits, which we refer to as a \textit{message block}; hence, unlike the former implementations, each element of the sequence corresponds to a block of bits and the corresponding transmitted channel input and the feedback symbols, which leads to a reduction in the sequence length by the message block size. Then, sequence-to-sequence encoding is performed on the sequence of message blocks and the corresponding knowledge vectors. Please see Fig. \ref{seq2seq} for a visualization of the sequence-to-sequence encoding scheme with message blocks. For sequence-to-sequence encoding, use of blocks of bits instead of single bits reduces the number of feedback iterations, under a fixed code rate, by the block size, which reduces the feedback overhead. Additionally, the achieved reduction on the length of the sequence also reduces the computational complexity and the memory requirements. 

Additionally, for the specific transformer-based DNN architecture that we will employ for our code design, when the elements of each sequence corresponds to a block of bits rather than a single bit, one can obtain a more informative embedding for each element of the sequence; and hence, improve the information processing capability of the transformers. This will become more clear when we introduce the details of the architecture below.

More formally, GBAF code design divides the $K$ original information bits into $l$ blocks of $m$ bits each. Here, we assume $m$ divides $K$, such that $K=l \cdot m$. These form our initial $l$ knowledge vectors (see Fig. \ref{seq2seq}). We utilize sequence-to-sequence encoding at each round of communication $\tau$, treating the $l$ knowledge vectors as the input sequence, and generate the symbols to be transmitted corresponding to each message block, and equivalently to each knowledge vector. Then, we update the sequence of knowledge vectors with the transmitted symbols and the received feedback symbols, by appending them to the corresponding knowledge vectors. A total of $N_{\tau}=l=K/m$ symbols, one parity symbol for each knowledge vector, are transmitted at each iteration $\tau$. One can observe that, given rate $R$ and block size $m$, the number of required communication rounds is $T=m/R$, which does not scale with $K$. Furthermore, by choosing different $T\in\mathbb{Z}^{+}$ and $m\in\mathbb{Z}^{+}$ values it is possible to obtain a wide range of code rate values $R=m/T$. Hence, the rate of the code can be adjusted by changing the block size $m$ and the number of communication rounds $T$, which is also equivalent to the total number of parity symbols transmitted per block. %We illustrate the encoding process sequence-to-sequence encoding with block formation utilized in GBAF code in Fig. \ref{seq2seq}. 
From the encoding process illustrated in Fig. \ref{seq2one} and Fig. \ref{seq2seq}, one can also observe that, under the same rate constraint, the sequence-to-sequence encoding approach requires $l$ times less interactions between the receiver and the transmitter compared to the sequence-to-one approach, which results in a reduced feedback overhead in practice, as argued above.

So far, we have identified two novel aspects of the GBAF code design; namely, utilizing sequence-to-sequence encoding instead of sequence-to-one encoding, and reorganizing the sequence before encoding by merging its elements into blocks of bits. Note that, these general design principles are independent from the particular DNN architecture that is used, and can be combined with any architecture that can be adapted for sequence-to-sequence encoding. The third novel aspect of the GBAF code is its architecture and the newly introduced modules. Different from the existing designs, we employ a novel \textit{transformer} encoder architecture for encoding and decoding. Furthermore, we introduce custom modules, such as a feature extractor to deal with large noise realizations. In order to provide a more holistic view from the design perspective, below we present the GBAF code design in three parts: the general architecture, the specific modules, and the implementation.

\subsection{GBAF Architecture}
From an operational point of view, we employ two types of components in the overall design, namely an encoder unit and a pre-processing unit. Motivated by the SK scheme \cite{SK1}, the transmitter consists of two cascaded units, each of which consists of a pre-processing unit followed by an encoder unit. We refer to the initial unit as the {\em belief network} and the latter as the {\em parity network}. The objective of the belief network is to generate a belief on the predicted bits at the receiver, while the objective of the {\em parity network} is to generate parity symbols to improve the prediction accuracy at the receiver. The receiver also employs a single unit with the same structure to predict the original bit stream, which we refer to as the decoder network. In the overall architecture, we identify three types of information flows, which we call as different  feedback mechanisms:
\begin{itemize}
    \item \textbf{Inner Feedback:} We use the term inner feedback to refer to a feedback mechanism within each unit. It is used for the encoder network to recall the previously generated parity symbols.
    \item \textbf{Belief Feedback:} The belief feedback refers to the information flow from the belief network to the encoder network.
    \item \textbf{Outer Feedback:} The outer feedback is the physical feedback signals from the receiver to the transmitter.
\end{itemize}

\begin{figure}[t]
\begin{center}
\includegraphics[scale =0.35]{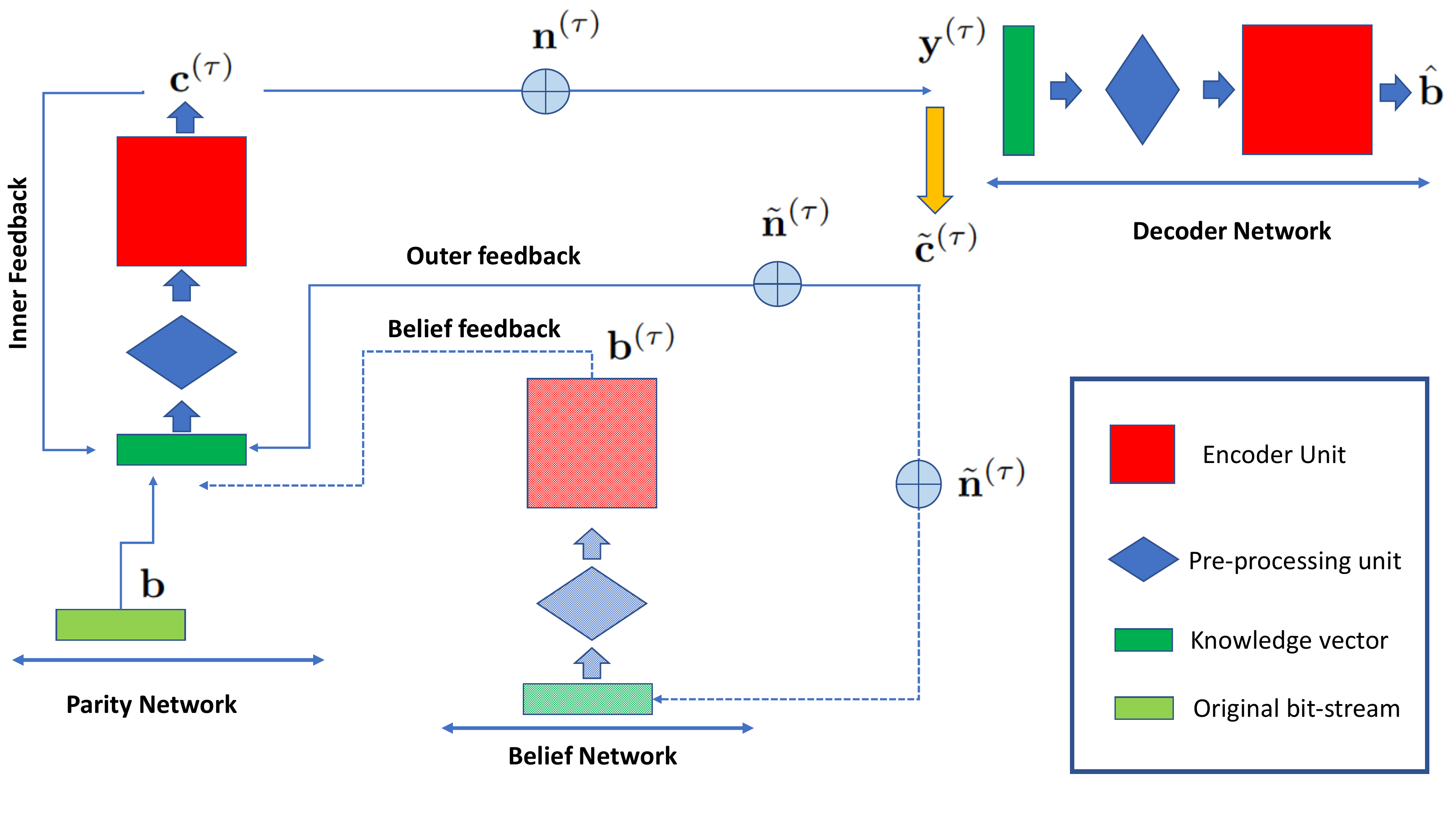}
\caption{Illustration of the overall GBAF code architecture. The green, blue and red blocks denote the knowledge vector, pre-processing unit, and encoder unit, respectively. The dashed lines and shapes indicate the units and connections that are optional.}
\label{sysmodel}
\end{center}
\end{figure}

The overall architecture is illustrated in Fig. \ref{sysmodel}. In the introduced architecture, the belief network and the belief feedback are optional; that is, they can be added or removed as desired, presenting a trade-off between complexity and performance, and the objective of using two networks is to disentangle the task of generating parity bits and predicting the belief at the receiver. However, by bypassing the belief network and disabling the belief feedback, both tasks can be fulfilled by the parity network.

\subsection{Modules}
\begin{figure}[t]
\begin{center}
\includegraphics[scale =0.4]{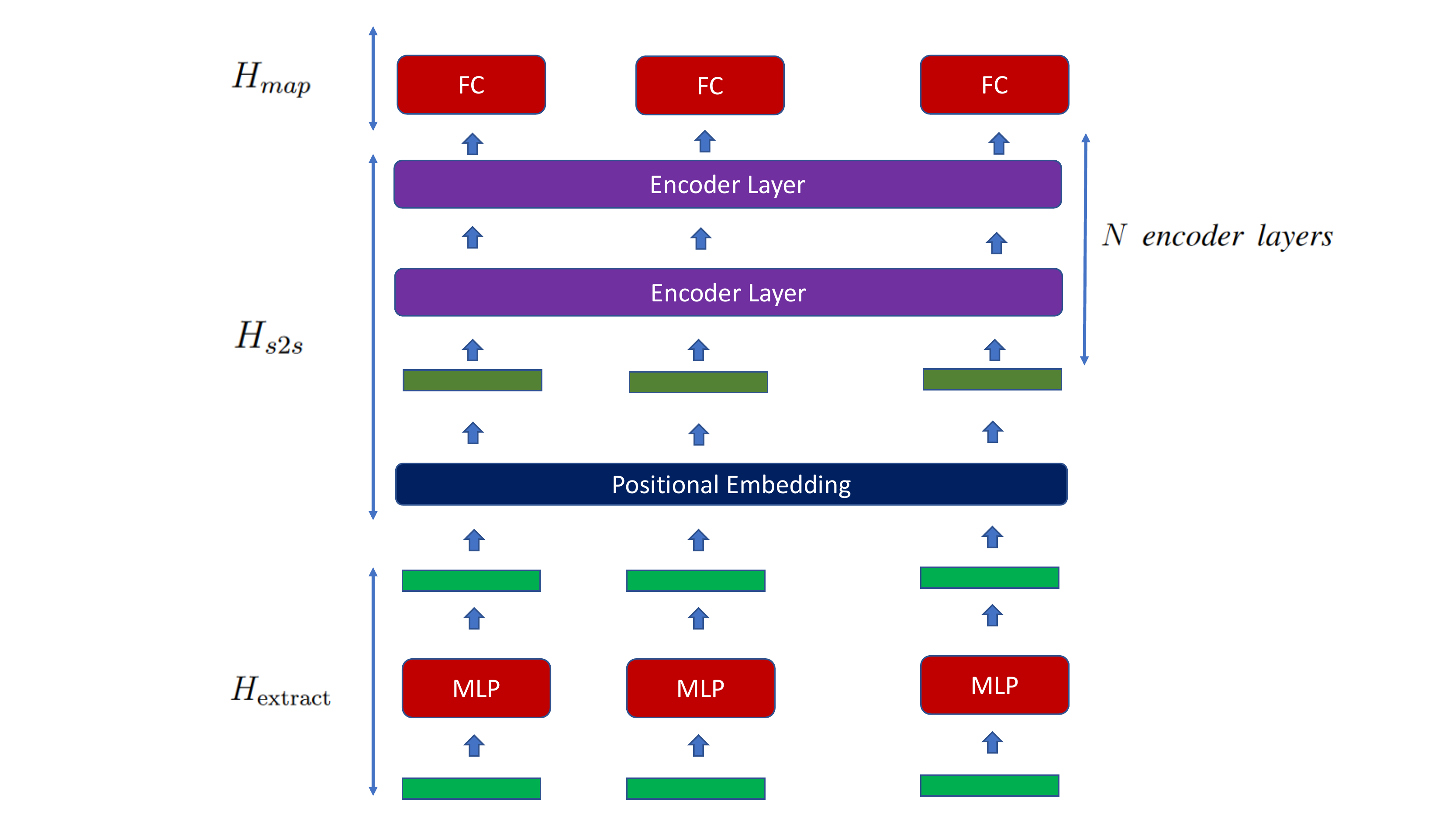}
\caption{Illustration of an encoder unit, $H_{\mathrm{encoder}}$.}
\label{encoder}
\end{center}
\end{figure}

In the GBAF code, for all the encoder units we utilize the same DNN architecture denoted by $H_{\mathrm{encoder}}$, which simply maps sequences of $l$ vectors of size $d_{in}$ to sequences of vectors of size $d_{out}$ with the same length, i.e., $H_{\mathrm{encoder}}( \bm{q}_{1},\ldots,\bm{q}_{l})= \mathcal{U}=\left\{\bm{u}_{1},\ldots,\bm{u}_{l} \right\}$,
such that $\bm{q}_{i}\in\mathbb{R}^{d_{in}}$, $\bm{u}_{i}\in\mathbb{R}^{d_{out}}$. $H_{\mathrm{encoder}}$ unit consists of three modules: feature extractor $H_{\mathrm{extract}}$, sequence-to-sequence encoder $H_{s2s}$, and output mapping $H_{map}$. Accordingly, $H_{\mathrm{encoder}} = H_{map} \circ H_{s2s} \circ H_\mathrm{extract}$, where $\circ$ denotes composition. The end-to-end architecture of the encoder unit is illustrated in Fig. \ref{encoder}. Below we explain each of these components in detail.

\subsubsection{Feature extractor}

The role of the feature extractor is to map the collected raw data for each block to a certain vector representation similar to the vector embedding approach in NLP tasks \cite{embed1, embed2, embed3}, where the objective is to represent the words in the form of a vector and the corresponding representation inherits certain contextual information regarding the word. However, our problem has two unique challenges: i) time-evolving nature of data; and ii) the randomness in the input. By the randomness, we refer to the random noise realization at each communication round. In principle, encoder module utilizes the noise realizations in the past to generate the parity symbols; nevertheless, the outlier noise realizations, particularly in the low SNR regime, might be overemphasized when a simple linear mapping is used for feature extraction. Hence, our ultimate aim is to design a feature extractor module in a way that the impact of each raw data on the corresponding representation is limited.
To this end, we utilize a multi-layer perceptron (MLP) architecture. As detailed in Appendix \ref{details}, the feature extractor consists of three linear layers with two activation functions in between. The activation function can be Gaussian error linear unit (GeLu) \cite{GeLu} or rectified linear unit (ReLu).  We use $H^{parity}_{extract}$ and $H^{belief}_{extract}$ to denote the feature extractors for parity network (line 14 of Algorithm \ref{alg:uipse}) and belief network (line 7 of Algorithm \ref{alg:uipse}), respectively.

\subsubsection{Sequence-to-sequence encoder}
Sequence-to-sequence encoder $H_{s2s}$ is a DNN architecture, where the sequence of early feature representations are mapped to a sequence of final latent representations by seeking certain correlations among the elements of the input sequence. The input to the $H_{s2s}$ is a sequence of $d_{model}$ dimensional vectors of length $l$ and the output is again a sequence of $d_{model}$ dimensional vectors of length $l$. Hence, a wide range of existing DNN architectures, particularly those employed for NLP, such as LSTM, GRU, transformer, can be utilized as $H_{s2s}$.  We use $H^{parity}_{s2s}$ and $H^{belief}_{s2s}$ to denote the sequence-to-sequence encoder for parity network (line 15 of Algorithm \ref{alg:uipse}) and belief network (line 8 of Algorithm \ref{alg:uipse}), respectively.

\begin{figure}[t]
\begin{center}
\includegraphics[scale =0.4]{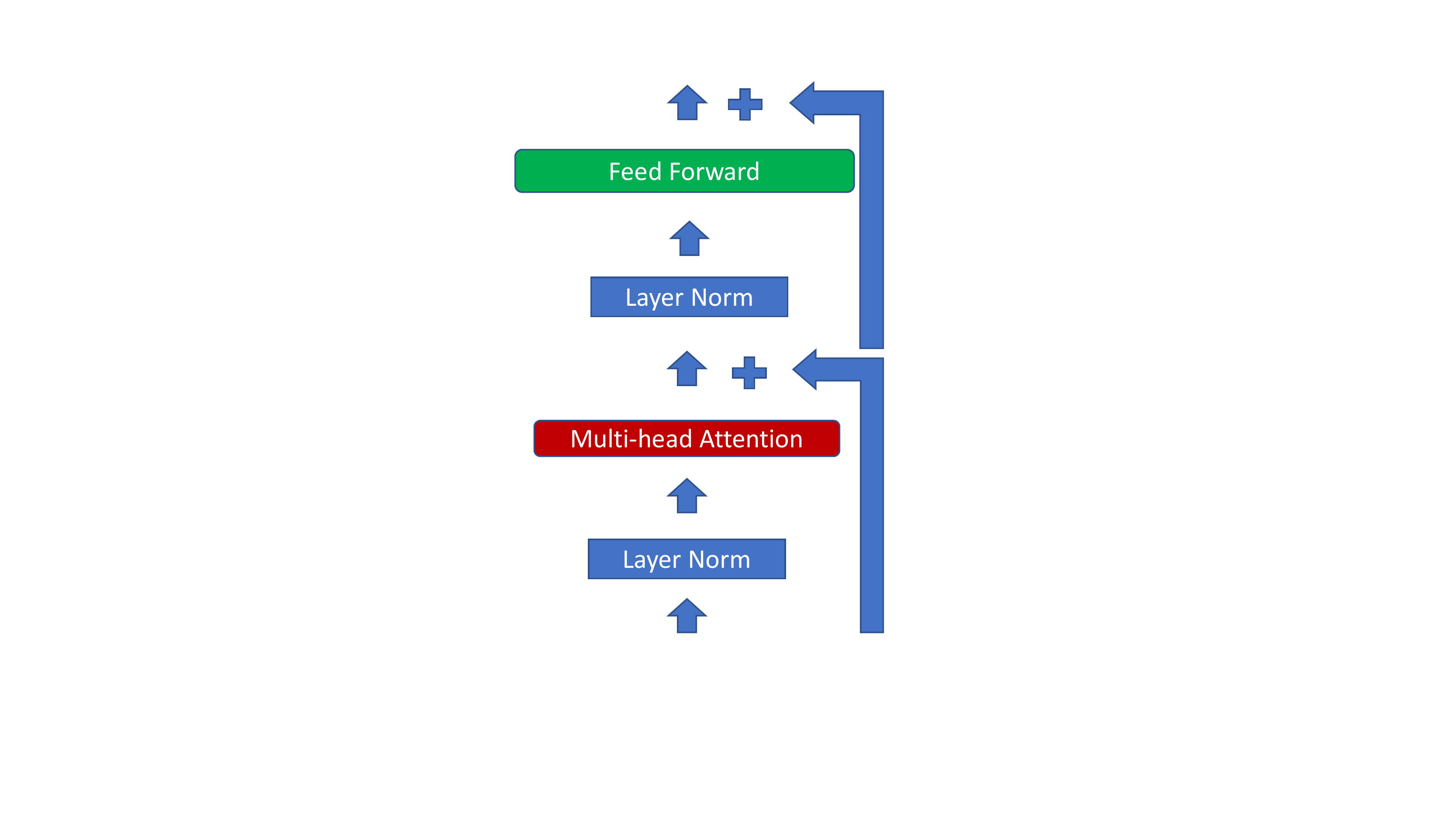}
\caption{Visualization of the encoder layer.}
\label{encoderlayer}
\end{center}
\end{figure}

We have observed that the transformer architecture performs particularly well for sequence-to-sequence encoding. Hence, for $H_{s2s}$, we consider sequence of $N$ {\em encoder layers} of the transformer architecture\footnote{We follow the standard implementation used in the Pytorch library:\\ \url{https://pytorch.org/docs/stable/_modules/torch/nn/modules/transformer.html#TransformerEncoderLayer}}, which consists of three main components: the feed forward module, the multi-head attention module, and the layer normalization module as illustrated in Fig. \ref{encoderlayer}. Next, we briefly explain the structure of the attention and feed forward modules.

\textbf{Attention Module:}
Attention mechanism is the key enabler of extracting relative information from a sequence. The core idea of the attention mechanism is to utilize a set of {\em key-value} pairs, for a given query, to generate an output. To be more precise, consider $d_{k}$, $d_{q}$, $d_{v}$ dimensional vectors of key, query and value, and a set of $N$ queries, a set of $K$ keys and a set of $K$ values, represented in a stacked form $\bm{Q}\in \mathbb{R}^{N\times d_{k}},\bm{K}\in \mathbb{R}^{K\times d_{k}},\bm{V}\in \mathbb{R}^{K\times d_{u}}$. The objective of the attention mechanism is to obtain weights required for combining the set of values to provide an output. The underlying mechanism used is scaled dot-product attention, i.e.,
\begin{equation}
Attn(\bm{Q},\bm{K},\bm{V})= \underbrace{Softmax\left(\frac{\bm{Q}\bm{K}^{T}}{\sqrt{d_{k}}}\right)}_{\bm{W}\in\mathbb{R}^{N\times K}}\bm{V},
\end{equation}
where ${\sqrt{d_{k}}}$ is used to normalized the output of the dot-product before the softmax layer. The multi-head attention follows the same principle, but query, key and  value vectors are first processed through a linear layer and fed to multiple attention mechanism/head, which are  executed simultaneously. Then the final output is obtained by concatenating the outputs of each attention mechanism/head. 

\textbf{Feed-Forward Module:}
The feed-forward module consists of two fully connected layers with a non-linearity (activation) between them, which can be  formally described as
\begin{equation}
FFN(\bm{x})=\phi(x\bm{W}+\bm{b})\tilde{\bm{W}} +\tilde{\bm{b}},
\end{equation}
where $\phi(\cdot)$ denotes a non-linear activation function, such as ReLU or GELU. Given an input vector of size $d_{model}$, first linear layer increases the model size to $\delta\times d_{model}$, which is later reduced to $d_{model}$ again by the second linear layer. Here, $\delta$ is often called the \textit{scaling factor}. It has been argued that the feed forward module functions as a memory \cite{ffn}. In our implementation, we set $d_{model}=32$, consider a single attention head, and set $\delta=4$ at the feed forward module following the common implementation \cite{attention2}, and finally for the layer normalization we follow the pre-layer normalization option \cite{normlayer}. For the number of encoder layers $N$, we consider $N_{parity}=2$, $N_{\mathrm{belief}}=2$ and $N_{\mathrm{decoder}}=3$. For further details about the transformer architecture we refer the reader to \cite{attention1, attention2, surveyat} and references therein.

\subsubsection{Output mapping}
The output mapping $H_{map}$ is used to map the final latent representation, obtained by the sequence-to-sequence encoder $H_{s2s}$, to a particular form depending on the purpose. For example, in the parity network, $H_{map}$ is used to map the final representation to a parity symbol, whereas in the belief network and decoder, it is used for classification purposes. The common aspect of $H_{map}$ in all three networks is that, it consists of a single fully-connected layer with an input size of $d_{model}$ and an output size of $d_{out}$; however, when it is used for classification, as in the belief and decoder networks, the fully-connected layer is followed by an additional softmax layer. 

Since only one parity symbol is generated per block, we consider $d_{out}=1$ for $H^{parity}_{map}$. On the other hand, decoder network aims to map each block to one of the $2^{m}$ possible $m$-length bit streams. Hence, for $H^{parity}_{map}$, we have $d_{out}=2^{m}$. On the other hand, for the belief network $H^{\mathrm{belief}}_{map}$, we set $d_{out}=2m$, that is for each original bit in the block we generate two values in order to represent the likelihood values $P(b_{i}=0)$ and $P(b_{i}=1)$ as a belief, with the help of a softmax layer\footnote{Here, we remark that before the softmax operation we reshape the input, i.e., $1\times2m \xrightarrow{} m\times2$.}. Finally, we note here that due to the average power constraint, an extra layer for power normalization is required following the $H^{parity}_{map}$, which follows the same procedure in \cite{deepcode, AttentionCode}.

\subsection{Implementation and Training Procedure}
Here, we illustrate how the proposed GBAF code architecture is executed from an algorithmic perspective in order to highlight its iterative structure. To describe the overall encoding procedure at the transmitter, we introduce an  iterative algorithm, called unified iterative parity symbol encoding (UIPSE), that generates $l$ symbols after each communication round, which is detailed in Algorithm \ref{alg:uipse}.

\begin{algorithm*}[t]
\caption{Unified iterative parity symbol encoding (UIPSE)}\label{alg:uipse}
\begin{algorithmic}[1]
\For{$\tau=1,\ldots,T$} $\#$ Generate 1 parity symbol per block at each pass
\State{\textbf{Update knowledge vector}:}
\State{$\boldsymbol{q}^{(\tau)}=[\bm{b},\bm{c}^{(1)},\ldots,\bm{c}^{(\tau)}, \tilde{\bm{y}}^{(1)},\ldots,\tilde{\bm{y}}^{(\tau-1)}]$}
\If{belief feedback is enabled}
\State{\textbf{Pre-process knowledge vector for belief network}:}
\State{$\left\{\tilde{\bm{q}}^{(\tau)}_{i},\ldots,\tilde{\bm{q}}^{(\tau)}_{l}\right\}=S_{\mathrm{belief}}(\boldsymbol{q}^{(\tau)})$,~~$\tilde{\bm{q}}^{(\tau)}_{i}=[\tilde{\bm{y}}^{(1)}_{i},\ldots,\tilde{\bm{y}}^{(\tau-1)}_{i}]$}
\State{\textbf{Extract features}:$\tilde{\bm{f}}^{(\tau)}_{i} = H^{\mathrm{belief}}_{extract}(\tilde{\bm{q}}^{(\tau)}_{i})$}
\State{\textbf{Attention-based neural-encoding}:$ \tilde{\mathcal{V}}^{(\tau)}= H^{\mathrm{belief}}_{\mathrm{s2s}}(\tilde{\mathcal{F}}^{(\tau)})$}
\State{\textbf{Generate belief feedback:}~~   $\bm{b}^{(\tau)}_{i}= H^{\mathrm{belief}}_{map}(\tilde{\bm{v}}^{(\tau)}_{i})$}
\State{\textbf{Pre-process knowledge vector}:$\left\{\bm{q}^{(\tau)}_{i},\ldots,\bm{q}^{(\tau)}_{l}\right\}=S_{parity}(\boldsymbol{q}^{(\tau)},\bm{b}^{(\tau)})$}
\Else
\State{\textbf{Pre-process knowledge vector}:$\left\{\bm{q}^{(\tau)}_{i},\ldots,\bm{q}^{(\tau)}_{l}\right\}=S_{parity}(\boldsymbol{q}^{(\tau)})$}
\EndIf
    \State{\textbf{Feature extraction:}}
    \For{$i\in[l]$}~~$\bm{f}^{(\tau)}_{i} = H^{parity}_{extract}(\bm{q}^{(\tau)}_{i})$
    \EndFor
    \State{\textbf{Attention-based neural-encoding}:$ \mathcal{V}^{(\tau)}= H^{parity}_{s2s}(\mathcal{F}^{(\tau)})$}
    \State{\textbf{Symbol mapping}:}
    \For{$i\in[l]$}
    \State{$c^{(\tau)}_{i}= H^{parity}_{map}(\bm{v}^{(\tau)}_{i})$} $\#$ Generate 1 parity symbol as feedback for $i$th block
    \EndFor
\EndFor
\end{algorithmic}
\end{algorithm*}

To describe the final decoding mechanism at the receiver, we introduce the joint parity symbol decoding (JPSD) algorithm, where the parity symbols belonging to each block are decoded jointly, as illustrated in Algorithm \ref{alg:decoder}. Different from the existing feedback code designs, due to the use of the block structure, the decoder performs classification over all possible bit blocks, $2^{m}$ in total, rather then binary classification. Hence, to recover the original bit-stream we further employ a lookup-table $\bm{A}$ (line 16-18 in Algorithm \ref{alg:decoder}), such that the $i$th row of $\bm{A}$, $\bm{A}_{[i,:]}$, corresponds to the bit-wise representation of the $i$th possible block.

From the training point of view, GBAF code performs a multi-class classification task. Let $\bm{x}\in\left\{0,1\right\}^m$ be the $m$-bit block to be transmitted. Then the relation between the data $\bm{x}$ and its label $y\in[0,\ldots,2^{m}-1]$ can be formulated as 
\begin{equation}\label{label}
y = \bm{x}^{T} \bm{z}
\end{equation}
where $\bm{z}= [2^{m-1},2^{m-2},\ldots,1]^{T}$. The data-label pairs $(\bm{x},y)$ are known at the transmitter. At the end of $T$ iterations, the receiver observes $\tilde{m}$-dimensional representation of $\bm{x}$, denoted by $\tilde{\bm{x}}$, and its task is to predict $y$ from observation $\tilde{\bm{x}}$.

To generate the training data, we first generate a random sequence of bits $\bm{b}\in\left\{0,1\right\}^{K}$, which is then divided into $l$ blocks, each of size $m$ bits, and assign the corresponding label for each block as described in (\ref{label}). Consequently, we have $\mathcal{B}=\left\{(\bm{b}_{1},y_{1}),\ldots,(\bm{b}_{l},y_{l})\right\}$ as the training data with corresponding labels. The generated block of bits, $\left\{\bm{b}_{1},\ldots,\bm{b}_{l}\right\}$ is then fed into the encoder, and at the end of $T$ communication iterations the decoder outputs a sequence of $l$ $2^{m}$-dimensional vectors,  $\left\{\bm{w}_{1},\ldots,\bm{w}_{l}\right\}$, as described in Algorithm \ref{alg:decoder}, which are then used to predict the class of each original block of bits. We use the cross-entropy loss function defined as 
\begin{equation}\label{loss}
L(\bm{W},Y) = \sum^{l}_{i=1}\sum^{2^{m}-1}_{c=0}-\log\frac{\exp{(\bm{W}_{[i,c]})}}{\sum^{2^{m}-1}_{c=0}\exp{(\bm{W}_{[i,c]})}}\cdot\mathbbm{1}_{y_{i}\neq c}
\end{equation}
where $Y=\left\{y_{1},\ldots,y_{l}\right\}$ denotes the labels of the blocks in the generated sequence, and $\bm{W}$ is the sequence $\left\{\bm{w}_{1}, \ldots, \bm{w}_{l}\right\}$ in matrix form. When a batch of sequences are generated for training, the loss function in (\ref{loss}) is evaluated by taking the average loss over the batch.

\begin{algorithm*}[t!]
\caption{Joint parity symbol Decoding (JPSD)}\label{alg:decoder}
\begin{algorithmic}[1]
\State{\textbf{Update Knowledge vector}:}
\State{$\hat{\boldsymbol{q}} =[\tilde{\bm{c}}^{(1)},\ldots,\tilde{\bm{c}}^{(T-1)}, \bm{y}^{(1)},\ldots,\bm{y}^{(T)}]$}
\State{\textbf{Pre-process knowledge vector for decoder network}:}
\State{$S_{\mathrm{decoder}}(\hat{\boldsymbol{q}})=\left\{\hat{\bm{q}}_{1},\ldots,\hat{\bm{q}}_{l}\right\}$,~~ $\hat{\bm{q}}_{i}=[\tilde{\bm{y}}^{(1)}_{i},\ldots,\tilde{\bm{y}}^{(T)}_{i}])$}
\State{\textbf{Feature extraction:}}
\For{$i\in[l]$}~~$\hat{\bm{f}}_{i} = H^{decoder}_{extract}(\hat{\bm{q}}_{i})$
\EndFor
\State{\textbf{Attention-based neural-encoding}:$ \hat{\mathcal{V}}= H^{decoder}_{s2s}(\hat{\mathcal{F}})$}
\State{\textbf{Mapping}:}
\For{$i\in[l]$}
$\bm{w}_{i}= H^{decoder}_{map}(\hat{\bm{v}}_{i})$
\EndFor
\State{\textbf{Block-wise classification}:} $\#$ predict the block index
\For{$i\in[l]$} $p_{i}= \max_{j}(\bm{w}_{i})_{[j]}$
\EndFor
\State{\textbf{Block index to bit stream conversion}:} $\#$ map block indecies to original bits
\For{$i\in[l]$} $\tilde{\bm{b}}=[\tilde{\bm{b}},\bm{A}_{[p_{i},:]}]$
\EndFor
\end{algorithmic}
\end{algorithm*}

\section{Numerical Results}\label{s:experiments}

In this section, we present the results of numerical experiments using the GBAF architecture and coding principles explained above.

\subsection{Experiment Setup}
In all the experiments, we consider a bit stream of length $K=51$ and a block size of $m=3$, which corresponds to $l=17$ blocks. We consider communication in the low forward SNR regime, where the availability of feedback can be particularly effective. Specifically, we consider $SNR_{ff}\in[-1,2]$~dB, and allow the transmission of $T=9$ parity bits for each block in total, which corresponds to a transmission rate of $R=3/9 = 1/3$.

For training, we utilized the AdamW optimizer, which is a variation of the Adam optimizer with decoupled weight decay regularization \cite{adamw}. It was observed in \cite{drf} that for DNN-aided code design, the training accuracy improves with the batch size. Accordingly, we consider a batch size of $B=8192$, the initial learning rate of $0.001$, and a weight decay parameter $0.01$. In addition, we apply gradient clipping with threshold $0.5$. We train the network for $100K$ batches using cross-entropy loss and apply polynomial decay to the learning rate.

Following the previous works, we consider the {block error rate (BLER)} as the performance measure for our analysis. We conduct our experiments under two different scenarios, noisy and noiseless feedback. In the noiseless feedback case, we use GeLu activation function in the feature extractor, while in the noisy feedback case, we use ReLu activation function. Further discussion on the impact of the activation function can be found in Appendix \ref{sec:GeLuReLu}.

% \begin{figure}[b]
%     \begin{subfigure}{0.47\textwidth}
%         \includegraphics[width=\linewidth]{fig_noiseless.pdf}
%         \caption{BLER performance versus the forward channel SNR (in dB). The feedback channel is noiseless.}
%         \label{Sim_noiseless}
%     \end{subfigure}
%     \hfill
%     \begin{subfigure}{0.47\textwidth}
%         \includegraphics[width=\linewidth]{FigS2.pdf}
%         \caption{BLER performance versus the forward channel SNR (in dB). The feedback channel SNR is $20$ dB.}
%         \label{Sim_noisy20}
%     \end{subfigure}
%     \caption{Performance comparison of GBAF with AttentionCode, DEFC, DRFC, DeepCode and NR-LDPC}
%     \label{sim}
% \end{figure}

\subsection{Experimental Results}
\begin{figure}
     \centering
     \begin{subfigure}
         \centering
         \includegraphics[width=0.48\columnwidth]{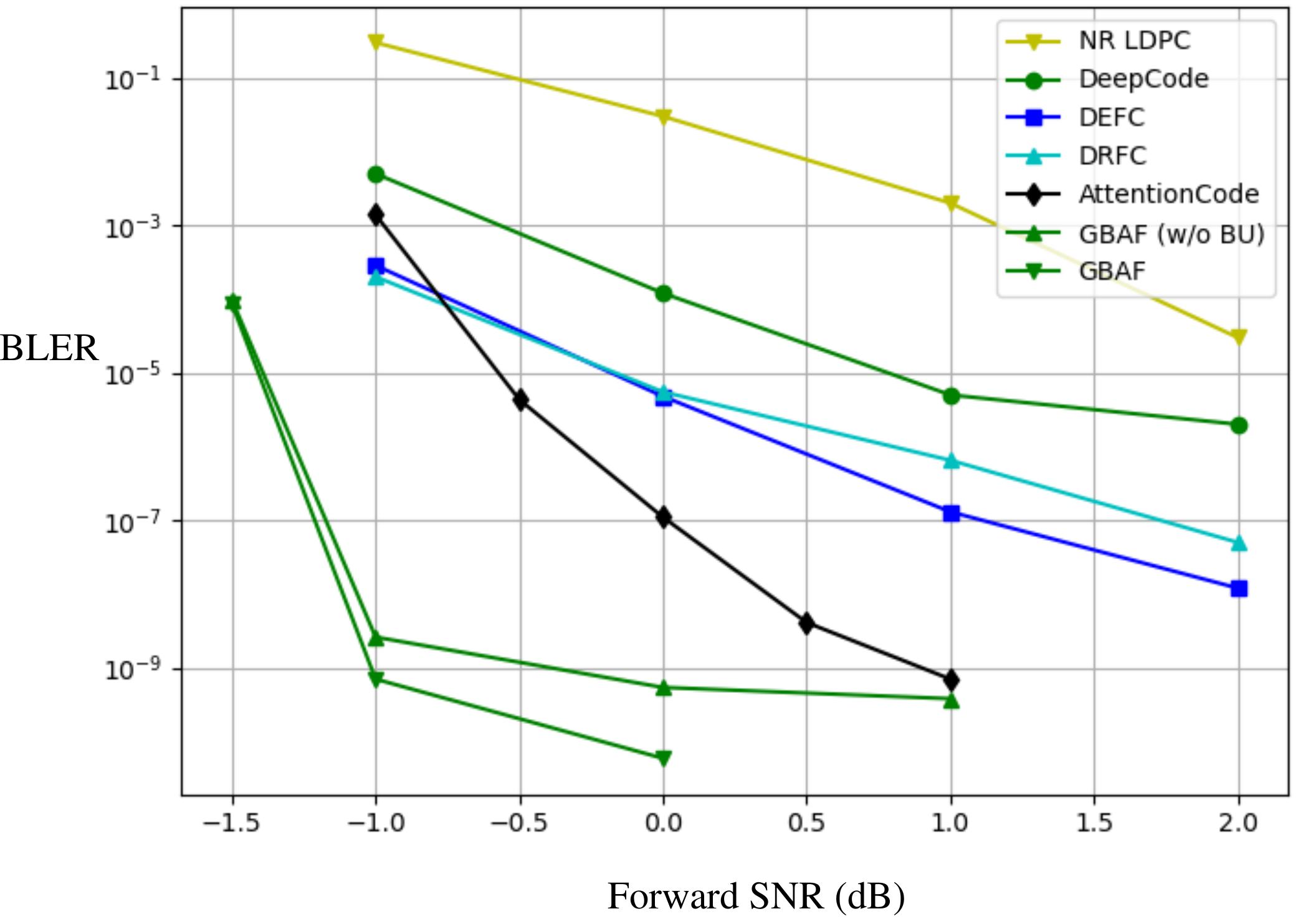}
     \end{subfigure}
     \begin{subfigure}
         \centering
         \includegraphics[width=0.48\columnwidth]{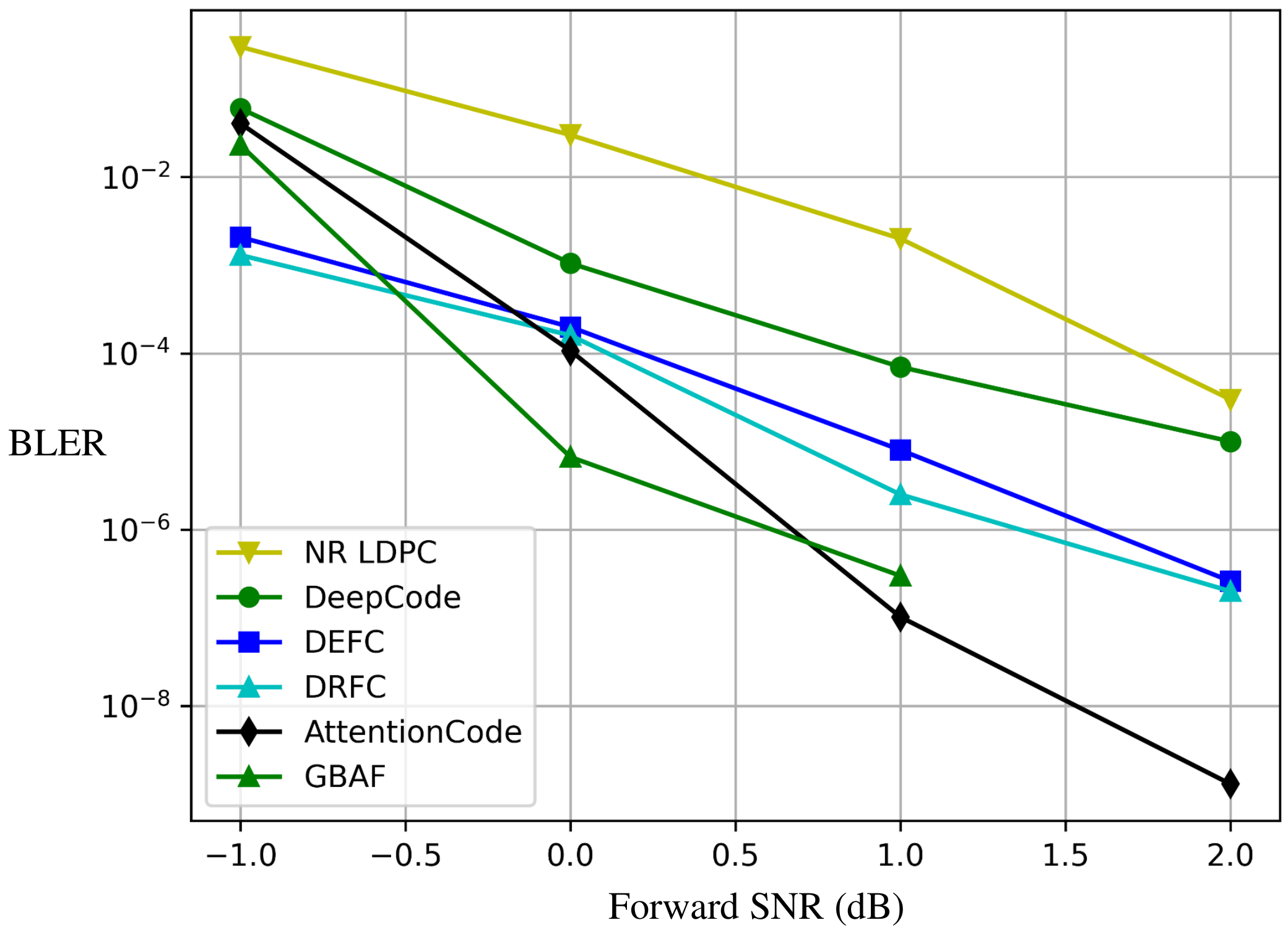}
     \end{subfigure}
    \caption{Performance comparison of GBAF with AttentionCode, DEFC, DRFC, DeepCode and NR-LDPC. GBAF (w/o BU) corresponds to the GBAF architecture without the belief network. (a) Noiseless feedback; (b) Noisy feedback.}
    \label{sim}
\end{figure}

We start our analysis with the noiseless feedback scenario, i.e., $\sigma^{2}_{fb}=0$. In the first part of the simulations, we focus on a fixed transmission rate of $R=3/9$, and compare the proposed design with the existing DNN-based feedback designs DeepCode\cite{deepcode}, DEFC\cite{defc}, DRFC\cite{drf}, AttentionCode\cite{AttentionCode} as well as LDPC code enhanced with neural decoder. In the first part of the simulations, we examine two variations of the GBAF code depending on the adoption of belief network in order to highlight its impact on the performance. The BLER performance results are illustrated in Fig. \ref{sim}(a) for the forward SNR values within the range of $[-1,2]$ dB. The results clearly highlight that GBAF code provides an order of magnitude improvement compared to the best performing alternative in the literature. We also observe that the adoption of the belief network further improves the performance. Nevertheless, design of the feature extraction module become more critical when the belief network is employed, and we observe that the belief network with the introduced feature extraction module may not be effective for the noisy feedback scenario. For now, we consider their joint design as an open research problem and in the remaining simulations we disable the belief network for the GBAF code.

Next, we consider the scenario in which the feedback channel is also exposed to additive Gaussian noise with $1/\sigma^{2}_{fb}=20$dB. The illustrated results in Fig \ref{sim}(b) indicate that except the lowest SNR value of $-1$dB, GBAF code outperforms  DEFC, DRFC, DeepCode and NR-LDPC. We also observe that at higher SNR values AttentionCode may outperform the GBAF code. However, we highlight the fact that GBAF code utilizes the feedback less frequently, approximately $6\times$ less, compared to the other codes considered in the figure. Hence, we can conclude that better or similar performance can be achieved with much less overhead. 

In the second set of simulations, we will highlight the flexibility of the proposed GBAF code design in terms of channel code rates it can achieve. Unlike the existing designs, the proposed framework can be easily adjusted to obtain codes at different rates by changing the number of parity symbols transmitted. This requires no variations in the architecture itself. To this end, we consider $T = 8, 7, 6, 5$ to achieve code rates $R = 3/8, 3/7, 3/6, 3/5$, respectively, and measure the BLER performance for forward SNR values in the range of $[-1,3]$ with noiseless feedback. The BLER performance achieved with these codes is presented in Table \ref{partrate}.

The results demonstrate that in the higher SNR regimes, it is possible to achieve acceptable BLER values with even higher code rates. For example, a BLER target of $10^{-5}$, which is sufficient for many tasks \cite{URLLC}, can be achieved at rates $R=3/8, 3/7, 3/6, 3/5$ for $SNR_{ff}=0, 1, 2, 3$ dB, respectively. Hence, unlike the existing designs, GBAF design exhibits certain  flexibility for rate adaptation based on the SNR.  We also notice from the table that the BLER performance degrades quickly as the code rate approaches the channel capacity at that SNR value. On the other hand, GBAF code manages to drastically lower the error rate when the code rate falls slightly below the capacity. We also note that the GBAF code performance in Fig. \ref{sim}(a) saturates to a BLER of $10^{-9}$ above $0$~dB; however, simulations at this BLER levels are less reliable as the code rarely observes any errors. Therefore, it is very unlikely to achieve BLER values lower than $10^{-9}$ even at higher SNRs. On the other hand, we can see in Table \ref{partrate} that, when the code rate is increased to $R=1/2$, the GBAF code performance does not saturate up until $3$~dB.

% \subsection{Limitations}
% {\color{red} One of the main limitations of the proposed GBAF architecture is that, due to the attention mechanism, the computational complexity of the architecture increases quadratically with the sequence length. Hence, the use of GBAF code for larger block sizes may prove difficult. On the other hand, there are recent works that achieve linear complexity in the attention mechanism \cite{linformer} without sacrificing the performance, and the adoption of a such light/sparse attention mechanism to the GBAF architecture for large blocklength code design can be considered as a future research direction.}

\subsection{Further Discussions}
The performance comparison between AttentionCode and GBAF code illustrates that the performance gain achieved by the GBAF code is not only related to the chosen sequence-to-sequence encoder architecture but also to the way it is implemented. Besides, compared to the AttentionCode implementation, GBAF code reduces the computational complexity and the memory requirement since the block encoding approach induces a reduction, linearly proportional to the block-size, in the sequence length. The computational complexity of the transformer architecture is  $\mathcal{O}(l^{2})$, although there are recent works targeting linear complexity \cite{reformer,linformer,performer}. This implies $m^{2}$ times reduction in complexity, which makes GBAF codes more practical compared to the AttentionCode for longer blocklengths.  On the other hand, the computational complexity still depends quadratically  on the sequence length, $l$. Hence, when the message length $K$ increases, limiting the complexity of the transformer architecture may require increasing the block size $m$. However, this would then significantly increase the complexity of the output mapping module $H^{parity}_{map}$, which grows exponentially as a function of $m$. Therefore, adoption of a light/sparse attention mechanism to the GBAF architecture for large blocklength code design, which would allow processing long sequences with limited computational complexity without sacrificing the performance, can be considered as an important future research direction.

We also remark that existing solutions, e.g., DeepCode or DRF code, transmit exactly two parity symbols at each communication round, thus for rate $R=3/9$, we need $T=52$ interactions between the forward and feedback channels, whereas GBAF requires only $T=9$, which implies a significant reduction in the overhead. Finally, we remark that by utilizing curriculum learning scheme used in \cite{AttentionCode}, the BLER performance can be improved further, especially for higher SNR values, where we observe certain saturation in the BLER performance.

\begin{table}[t]
\centering
\caption{BLER of GBAF codes with different code rates $R$.}
\label{partrate}
\begin{tabular}{@{}lllll@{}}
\toprule
\textbf{SNR/Rate} &  3/8  &  3/7 &  3/6 & 3/5 \\ \midrule
-1 dB & $1.8\times10^{-2}$ & - & - & - \\
0 dB & $6.15\times10^{-8}$ & $2.8\times10^{-3}$ & - & - \\
1 dB & $2.7\times10^{-8}$& $7.5\times10^{-8}$ & $1\times10^{-2}$& - \\ 
2 dB & - & $1\times10^{-9}$ & $1.5\times10^{-6}$ & $6.5\times10^{-2}$\\ 
3 dB & - & - & $2.7\times10^{-8}$& $8.7 \times10^{-7}$\\  \bottomrule
\end{tabular}
\end{table}

\subsection{Fading channels}
In this section, we evaluate the performance of GBAF codes over fading channels. 
In particular, we consider the fading channel defined in new radio (NR) clustered delay line (CDL).
A CDL is used to model the channel when the received signal consists of multiple delayed clusters, where each cluster contains multi-path components with the same delay but slight variations in the angles of departure and arrival.

We consider communication between a mobile user and the gNodeB (gNB) with GBAF code, where the mobile user is node A and the gNB is node B. The speed of the mobile user is $v_u$ m/s and the root-mean-square (RMS) delay spread is $100$ ns.
The 5G system is configured with a carrier frequency of $3.5$ GHz, a subcarrier spacing of $30$ KHz, and a slot duration of $0.5$ ms.

\begin{figure}[t]
  \centering
  \includegraphics[width=0.4\columnwidth]{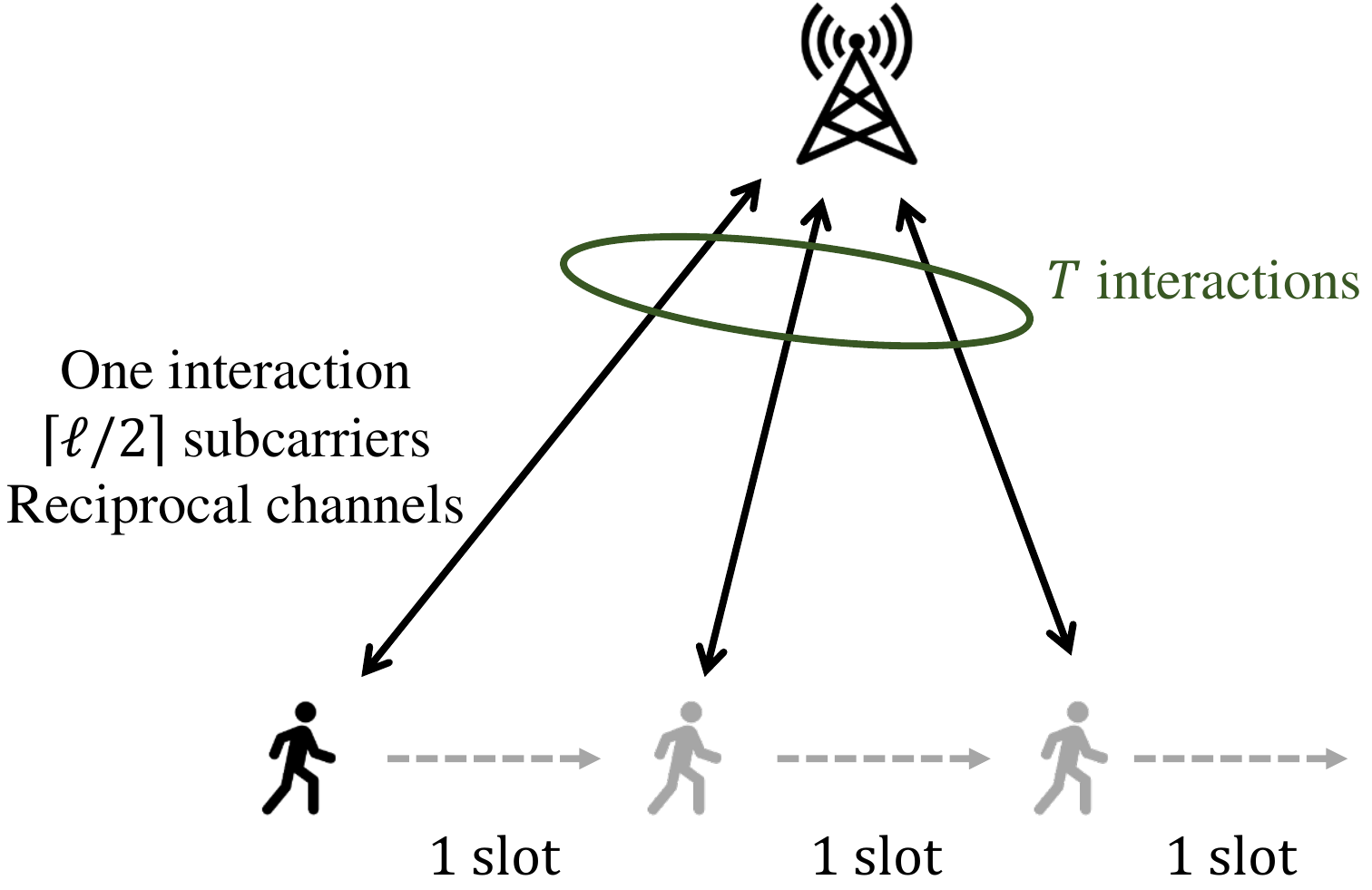}\\
  \caption{The communication between a mobile user and a GNB with GBAF code. The communication lasts for $T$ interactions, corresponding to $T$ slots. In each interaction, the $\ell$ codes symbols are transmitted from the mobile user to the gNB (feedforward link) in $\lceil \ell/2 \rceil$ subcarriers; the feedback is transmitted from the gNB to the mobile user (feedback link) via the same $\lceil \ell/2 \rceil$ subcarriers, where we assume reciprocal channels in one interaction.}
\label{fig:fadingmodel}
\end{figure}

As shown in Fig.~\ref{fig:fadingmodel}, the user communicates with the gNB in $T$ interactions. One interaction corresponds to one slot and the $\ell$ real coded symbols are modulated onto $\lceil \ell/2 \rceil$ subcarriers. We assume reciprocal channels, meaning that the channel gains of the $\lceil \ell/2 \rceil$ subcarriers are the same for the uplink (from the user to the gNB) and downlink (from the gNB to the user) transmissions in one interaction.

\subsubsection{Channel-gain generation}
Channel gains are generated by QUAsi Deterministic RadIo channel GenerAtor (QuaDRiGa) \cite{QuaDRiGa} using the CDL-Model for NLOS (3GPP TR38.901 NR-CDL-C). Specifically, for a given mobile speed $v_u$, we generate two long move paths of the mobile user and record the channel-gain variations of the $\lceil \ell/2 \rceil$ subcarriers. Let us denote the two long trajectories of channel-gains by $\text{Tr}_t(v_u)$ and $\text{Tr}_e(v_u)$, respectively. $\text{Tr}_t(v_u)$ will be used in the training phase and $\text{Tr}_e(v_u)$ will be used in the evaluation phase.
In the training phase, we randomly sample an initial point in $\text{Tr}_t(v_u)$ and extract the channels gains of $T$ consecutive slots starting from the initial point as the channel gains of one training epoch (one training epoch means one communication round to deliver a stream of $K$ bits). Likewise, in the evaluation phase, we randomly sample an initial point in $\text{Tr}_e(v_u)$ and extract the channels gains of $T$ consecutive slots starting from the initial point as the channel gains of one evaluation epoch to assess the performance of a well-trained GBAF code model.

\subsubsection{Sample statistics of the channel gains}
Next, we analyze the sample statistics of the channel gains in $\text{Tr}_t(v_u)$ and $\text{Tr}_e(v_u)$. Two main results are as follows:
\begin{itemize}
    \item Channel gains are correlated across subcarriers and interactions: across subcarriers, the fading is almost flat; across interactions, the amplitude and phase of the channel gains progressively increase or decrease.
    \item For each subcarrier, the statistical properties of the channels are almost the same. Let us focus on one subcarrier and analyze the statistical properties of its amplitude and phase.
\begin{figure}[t]
  \centering
  \includegraphics[width=0.8\columnwidth]{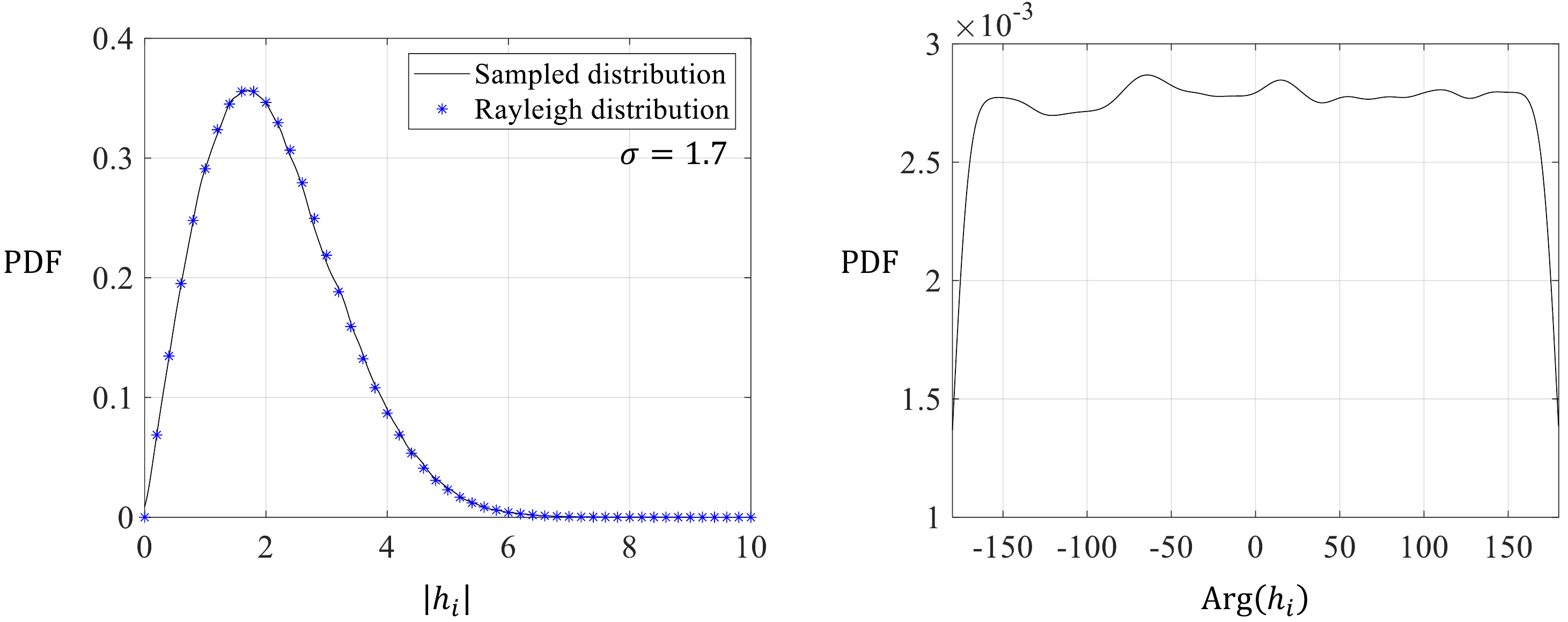}\\
  \caption{The PDF of the amplitude and phase obtained from the sampled channel gains of a subcarrier.}
\label{fig:dist}
\end{figure}

Let $v_u=1$ m/s and consider the $i$-th subcarrier, $i=1,2,...,\lceil \ell/2 \rceil$. Fig.~\ref{fig:dist} presents the probability density function (PDF) of $|h_i|$ and $\text{Arg}(h_i)$.
As can be seen, the PDF of $|h_i|$ can be fitted by a Rayleigh distribution with $\sigma=1.7$. The PDF of $\text{Arg}(h_i)$, on the other hand, is approximately a uniform distribution. This indicates that the channel coefficients generated by QuaDRiGa for a single subcarrier can be viewed as Rayleigh fading.
\end{itemize}

\subsubsection{Adapting GBAF code to fading channels}
To apply GBAF code in fading channels, we assume that the channel gains are perfectly known to both nodes A (mobile user) and B (gNB).
The received symbols at nodes A and B in fading channels can be written as 
\begin{eqnarray}
\label{eq:yfading1}
&&\hspace{-0.5cm} \bm{y} = \bm{h}\odot\bm{c} + \bm{n}, \\
\label{eq:yfading2}
&&\hspace{-0.5cm} \widetilde{\bm{y}} = \widetilde{\bm{h}}\odot\widetilde{\bm{c}} +\widetilde{\bm{n}}= \widetilde{\bm{h}}\odot \bm{h}\odot\bm{c} +\widetilde{\bm{h}}\odot \bm{n} +\widetilde{\bm{n}},
\end{eqnarray}
where $\odot$ denotes the element-wise product; $\bm{h}$ and $\widetilde{\bm{h}}$ are the feedforward and feedback channel gains, respectively. In the case of reciprocal channels, we have $\bm{h}=\widetilde{\bm{h}}$.

Given the knowledge of $\bm{h}$, nodes A and B transform \eqref{eq:yfading1} and \eqref{eq:yfading2} to 
\begin{eqnarray}
\label{eq:yfading3}
&&\hspace{-0.5cm} \bm{y}\odot \frac{1}{\bm{h}} = \bm{c} + \bm{n}\odot \frac{1}{\bm{h}}, \\
\label{eq:yfading4}
&&\hspace{-0.5cm} \widetilde{\bm{y}}\odot \frac{1}{\bm{h}}\odot \frac{1}{\bm{h}} = \bm{c} + \bm{n} \odot \frac{1}{\bm{h}} +\widetilde{\bm{n}} \odot \frac{1}{\bm{h}} \odot \frac{1}{\bm{h}}.
\end{eqnarray}
In so doing, the fading coefficients are transformed into the noise terms -- the architecture of GBAF code can be used with the only difference being the non-AWGN noise.

\begin{figure}[t]
  \centering
  \includegraphics[width=0.8\columnwidth]{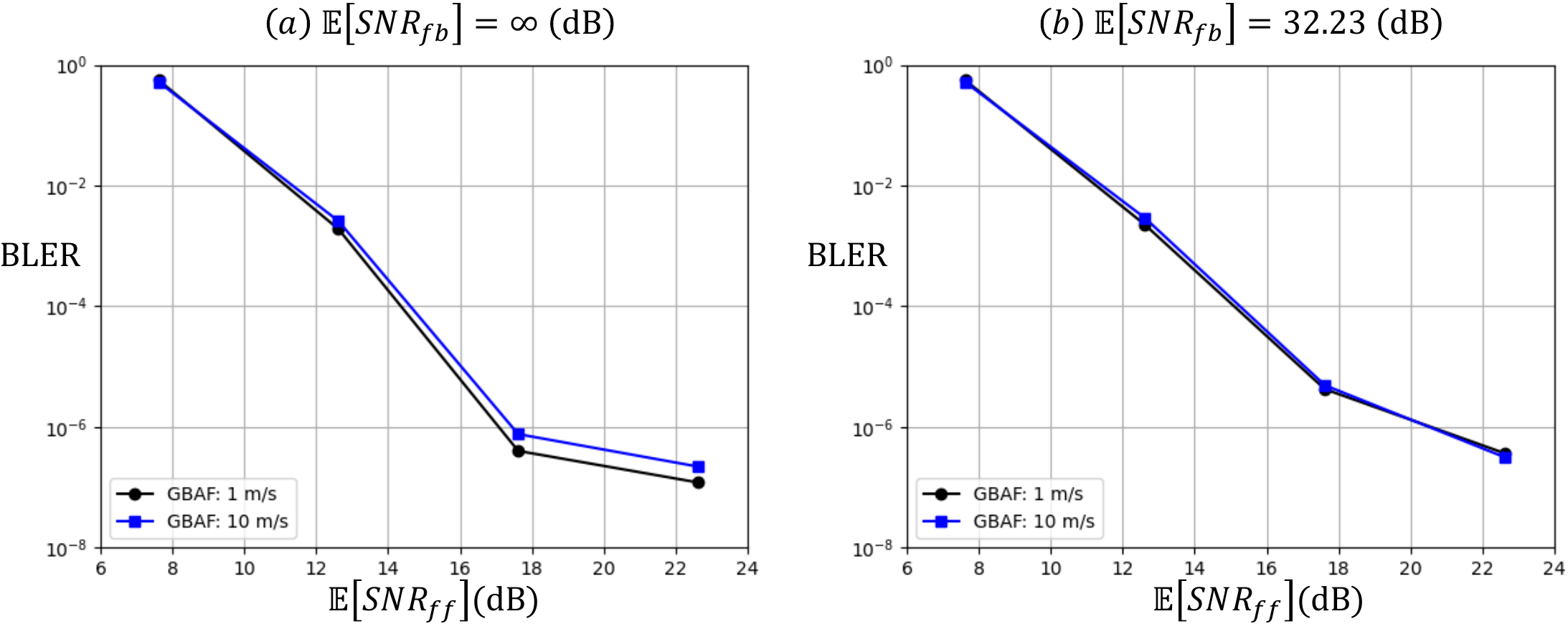}\\
  \caption{Performance of GBAF code in fading channels benchmarked against DeepCode.}
\label{fig:sim_fading}
\end{figure}

\subsubsection{Performance evaluation}
Under the above setup, this subsection evaluates the performance of GBAF codes in fading channels. In the simulations, we generate the channel-gain trajectories using two user speeds $v_u=1$ m/s and $10$ m/s. The feedback SNR depends on both fading coefficients and noise power. When the feedback channel is noiseless (noise power is $0$), the feedback SNR is simply $SNR_{fb}=\infty$ dB. When the feedback channel is noisy, we fix the average feedback SNR to $\mathbb{E}[SNR_{fb}]=32.23$ dB (in which case the noise power is the same as that of the AWGN channel case with $20$ dB noisy feedback).

Fig.~\ref{fig:sim_fading} presents the BLER of GBAF code, where the feedback channel is noiseless in (a) and noisy in (b).
Note that we do not simulate any prior works as benchmarks because they are designed exclusively for the unit-time delay case and do not fit into the considered NR-CDL model.
As shown in Fig.~\ref{fig:sim_fading}, although faster mobile speed leads to faster-changing channel gains, GBAF code is robust to the mobile speed. The BLER performance only degrades slightly when the mobile speed increases from 1 m/s to 10 m/s.

\section{Conclusion}\label{s:Conclusion}
In this work, we have introduced the novel generalized block attention feedback (GBAF) codes, which is empowered by the sequence-to-sequence encoding DNN architecture, particularly the transformer architecture, to generate parity bits, by incorporating a feedback mechanism. Beyond introducing a generic framework, unlike the existing solutions described through the employed DNN architecture, the proposed framework also addresses several practical limitations of the existing DNN-based coding approaches and makes DNN-based feedback codes more applicable for next generation networks. In particular, our architecture is not limited to a fixed code rate, and achieves a significantly lower overhead thanks to its block structure. Finally, in addition to these operational advantages, we have also shown that GBAF codes significantly outperform the existing solutions, especially in the noiseless feedback scenario. This will be particularly attractive for applications where the feedback link is from a base station to a user equipment, and hence, can be assumed to achieve relatively high SNR values. We also showed that GBAF codes can be robust to channel fading, making them a promising alternative for mobile channels with relatively good average channel conditions.

\appendices

\section{Implementation Details}\label{details}

%\subsection{Details on the pre-processing unit}
The pre-processing unit of the encoder network can be operated under four different mode based on the enabled/disabled feedback mechanisms and the way the available information are aggregated. In Algorithm \ref{alg:pre-process}, we illustrate pre-processing mechanism under each mode with different color. In our implementation, we prefer the third option illustrated with {\color{blue} blue}. Here, we also note that in our implementation we store the original bits $\bm{b}$ in the knowledge vector in the BPSK modulated form, i.e., $\bar{\bm{b}}=2*\bm{b}-1$. In overall, with enable/disable option for the belief network the GBAF code design can be operated under 8 different modes as highlighted in Algorithm \ref{alg:pre-process}.

\begin{algorithm*}[h!]
\caption{Pre-processing unit for  Parity Network: $S_{parity}()$}\label{alg:pre-process}
\begin{algorithmic}[1]
\If{Belief $\bm{b}^{(\tau)}$ is available}
\If{Feedback only is \textbf{True}}
\State{{\color{red}$\bm{q}^{(\tau)}_{i}=[\bm{b}_{\left((i-1)*m+1:i*m\right)},\bm{b}^{(\tau)}_{i},\tilde{\bm{y}}^{(1)}_{i},\ldots,\tilde{\bm{y}}^{(\tau-1)}_{i}])$}}
\ElsIf{Noise only is \textbf{True}}
\State{{\color{orange}$\bm{q}^{(\tau)}_{i}=[\bm{b}_{\left((i-1)*m+1:i*m\right)},\bm{b}^{(\tau)}_{i},\tilde{\bm{y}}^{(1)}_{i}-\bm{c}^{(1)}_{i},\ldots,\tilde{\bm{y}}^{(\tau-1)}_{i}- \bm{c}^{(\tau-1)}_{i}])$}}
\ElsIf{Disentangle is \textbf{True}}
    \State{{\color{blue}$\bm{q}^{(\tau)}_{i}=[\bm{b}_{\left((i-1)*m+1:i*m\right)},\bm{b}^{(\tau)}_{i}, \bm{c}^{(1)}_{i},\ldots,\bm{c}^{(\tau-1)}_{i}, \tilde{\bm{y}}^{(1)}_{i}-\bm{c}^{(1)}_{i},\ldots,\tilde{\bm{y}}^{(\tau-1)}_{i}- \bm{c}^{(\tau-1)}_{i}]$}}
    \Else
    \State{{\color{green}$\bm{q}^{(\tau)}_{i}=[\bm{b}_{\left((i-1)*m+1:i*m\right)},\bm{b}^{(\tau)}_{i}, \bm{c}^{(1)}_{i},\ldots,\bm{c}^{(\tau-1)}_{i}, \tilde{\bm{y}}^{(1)}_{i},\ldots,\tilde{\bm{y}}^{(\tau-1)}_{i}]$}}
    \EndIf
    \Else 
 \If{Feedback only is \textbf{True}}
\State{{\color{red}$\bm{q}^{(\tau)}_{i}=[\bm{b}_{\left((i-1)*m+1:i*m\right)},\tilde{\bm{y}}^{(1)}_{i},\ldots,\tilde{\bm{y}}^{(\tau-1)}_{i}])$}}
\ElsIf{Noise only is \textbf{True}}
\State{{\color{orange}$\bm{q}^{(\tau)}_{i}=[\bm{b}_{\left((i-1)*m+1:i*m\right)},\tilde{\bm{y}}^{(1)}_{i}-\bm{c}^{(1)}_{i},\ldots,\tilde{\bm{y}}^{(\tau-1)}_{i}- \bm{c}^{(\tau-1)}_{i}])$}}
\ElsIf{Disentangle is \textbf{True}}
    \State{{\color{blue}$\bm{q}^{(\tau)}_{i}=[\bm{b}_{\left((i-1)*m+1:i*m\right)},\bm{c}^{(1)}_{i},\ldots,\bm{c}^{(\tau-1)}_{i}, \tilde{\bm{y}}^{(1)}_{i}-\bm{c}^{(1)}_{i},\ldots,\tilde{\bm{y}}^{(\tau-1)}_{i}- \bm{c}^{(\tau-1)}_{i}]$}}
    \Else
    \State{{\color{green}$\bm{q}^{(\tau)}_{i}=[\bm{b}_{\left((i-1)*m+1:i*m\right)}, \bm{c}^{(1)}_{i},\ldots,\bm{c}^{(\tau-1)}_{i}, \tilde{\bm{y}}^{(1)}_{i},\ldots,\tilde{\bm{y}}^{(\tau-1)}_{i}]$}}   
\EndIf
\EndIf
\end{algorithmic}
\end{algorithm*}

\section{Feature extractors}\label{sec:GeLuReLu}
The feature extractor presented in the main body of this paper is selected from a bunch of different designs. In this appendix, we discuss these designs and explain how the feature extractor is chosen.

\subsection{Various designs of the feature extractor}\label{extractors}
\begin{figure}[ht]
     \centering
     \begin{subfigure}
         \centering
         \includegraphics[width=0.9\columnwidth]{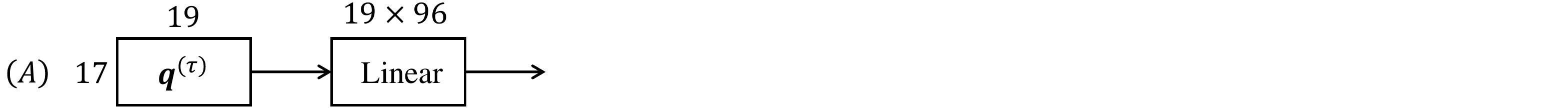}
     \end{subfigure}
     \hfill
     \begin{subfigure}
         \centering
         \includegraphics[width=0.9\columnwidth]{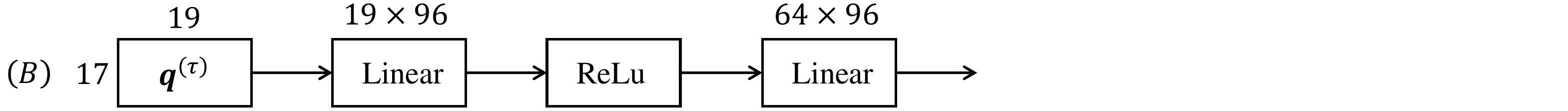}
     \end{subfigure}
     \hfill
     \begin{subfigure}
         \centering
         \includegraphics[width=0.9\columnwidth]{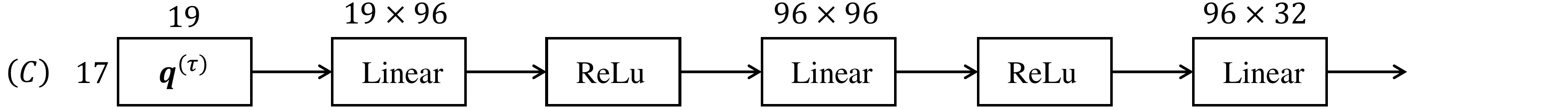}
     \end{subfigure}
     \hfill
     \begin{subfigure}
         \centering
         \includegraphics[width=0.9\columnwidth]{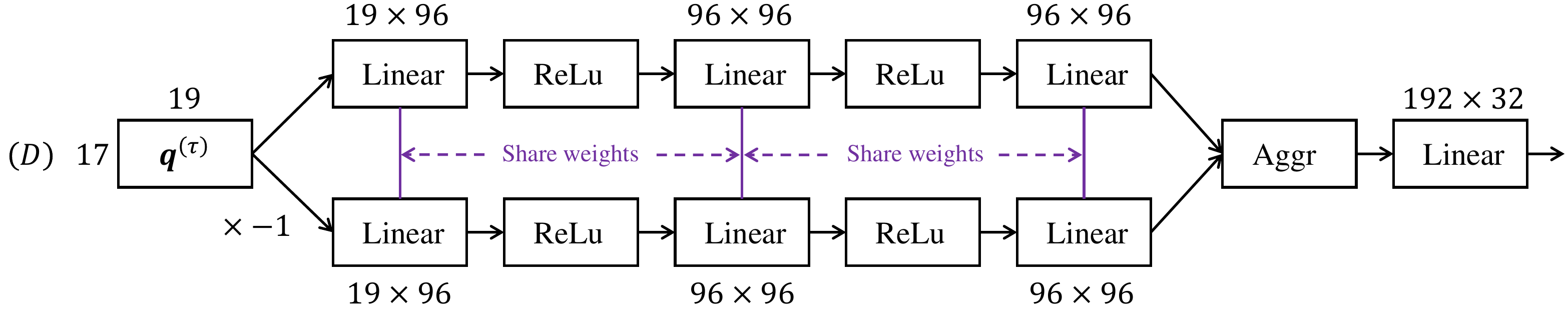}
     \end{subfigure}
     \hfill
     \begin{subfigure}
         \centering
         \includegraphics[width=0.9\columnwidth]{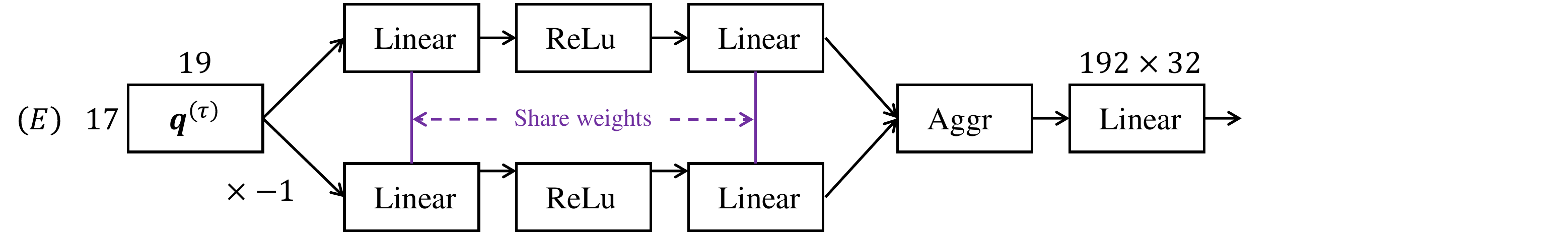}
     \end{subfigure}
     \hfill
     \begin{subfigure}
         \centering
         \includegraphics[width=0.9\columnwidth]{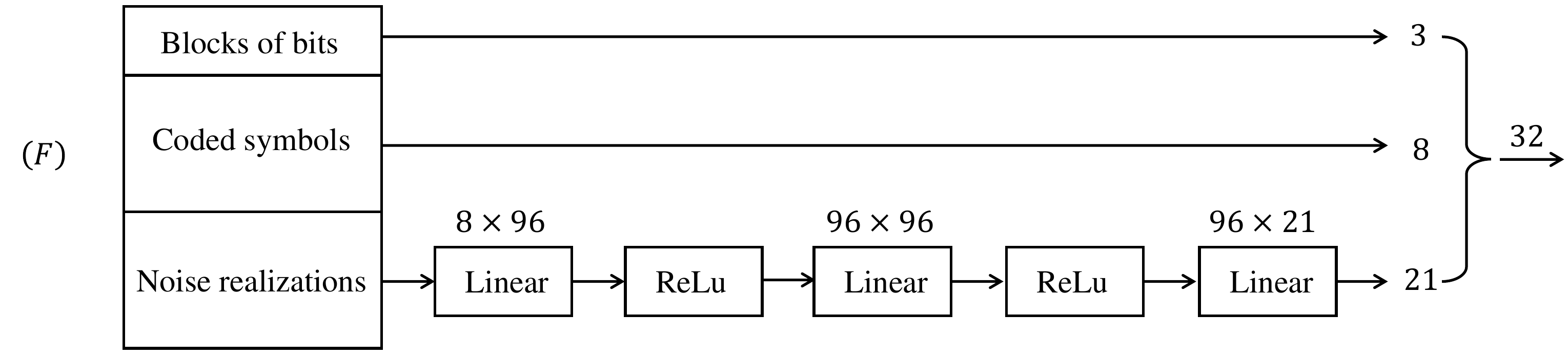}
     \end{subfigure}
     \hfill
     \begin{subfigure}
         \centering
         \includegraphics[width=0.9\columnwidth]{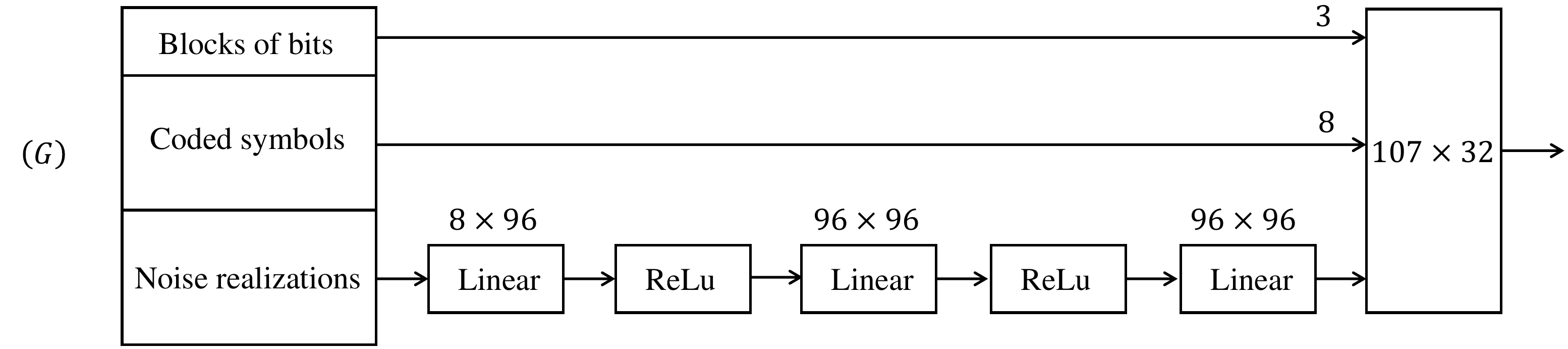}
     \end{subfigure}
        \caption{Seven designs of the feature extractor.}
        \label{fig:designs}
\end{figure}

We propose seven different designs of feature extractor for GBAF code, the architectures of which are summarized in Fig.~\ref{fig:designs}.
Note that variations of these designs can be obtained by changing the number of neurons in each layer or the activation functions.
\begin{itemize}
\item Design A is simply a linear layer. This is the feature extractor used in the original design of transformer \cite{attention1}.
\item Design B consists of two linear layers with a ReLu activation function in between. The output of ReLu is $0$ when the input is negative. Therefore, it can be used for truncation: whenever the DNN wants to truncate a large output of a neuron (large in amplitude), say $z$, it can simply multiply $z$ by a weight $-1$ (when $z>0$) or $1$ (when $z<0$) and then feed $-z$ or $z$ into ReLu, yielding $\text{ReLu}(-|z|) = 0$.

% \begin{figure}[h]
%   \centering
%   \includegraphics[width=0.5\columnwidth]{figures/GeLu.pdf}\\
%   \caption{The non-linear activation functions: ReLu, ELu, GeLu \cite{GeLu}.}
% \label{fig:gelu}
% \end{figure}

With design B, we hope the ReLu non-linearity can truncate noise realizations with large amplitude, mimicking the modulo operation used in modulo-SK \cite{ModuloSK}. It is worth noting that ReLu can be replaced by GeLu, which will be discussed later.

\item Design C is an extension of design B, where we use three linear layers with two ReLu activation functions in between.  A single truncation layer in design B can only truncate either positive or negative noise realizations when the weights and bias of the linear layers are fixed. This motivates us to add an additional ReLu truncation in Design C such that one ReLu can truncate positive  noise realizations and the other can truncate negative noise realizations.

Design C is the final design we choose as the default feature extractor for GBAF code.

\item In Design D, we use two parallel noise suppression flows and each flow is the same as design C. In particular, 1) for the second flow, the noise realization part is multiplied by $-1$ as the input; 2) the two  parallel flows share the same weights. In doing so, we specifically ask the DNN to tackle the positive noise realization in one flow and the negative noise realization in the other flow. After processing by two parallel branches, the resulting features are aggregated and transformed by a linear layer to obtain the output.

\item Design E is a simplification of design D, where each of the two parallel flows is chosen to be design B, as opposed to design C.

\item Designs F and G are extensions of design C. Specifically, instead of processing the whole feature matrix, we only process the noise realizations by design C. After feature extraction, we aggregate the extracted features with bits and coded symbols to obtain the final features to be fed into the attention network. The difference between designs F and G lies in the last aggregation step: design F simple aggregates the features, while the aggregation in design G is followed by a linear transformation.
\end{itemize}

\begin{figure}[t]
  \centering
  \includegraphics[width=0.7\columnwidth]{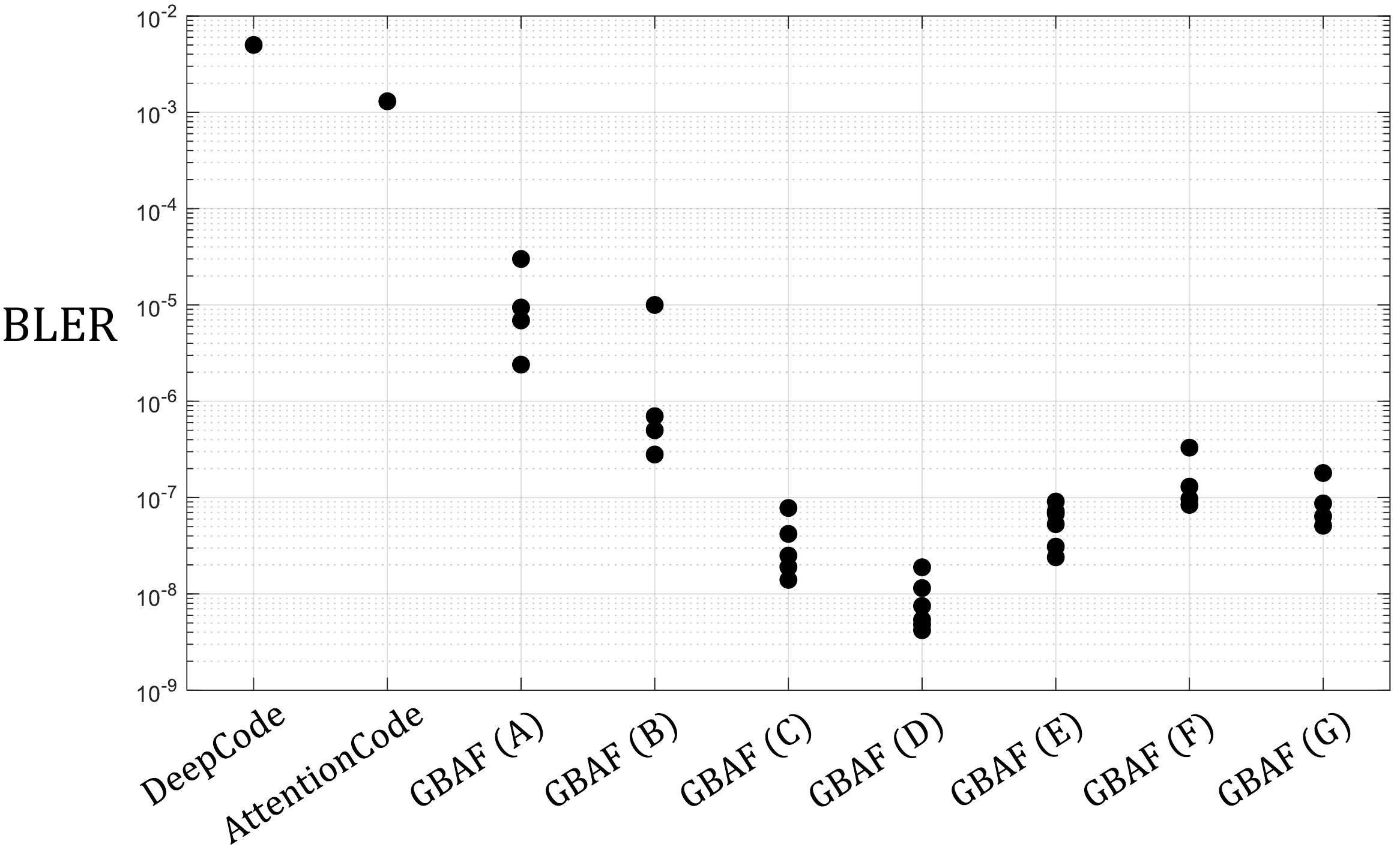}\\
  \caption{BLER performance of various designs. The system setup is $K=51$, $N = 153$, $m=3$, $\ell=17$, the feedforward SNR is fixed to $-1$ dB, and the feedback is noiseless. One black point stands for one independent simulations.}
\label{fig:S1}
\end{figure}

\subsection{Architecture selection}
This section compares the seven designs in Fig.~\ref{fig:designs} considering the noiseless feedback system. In particular, we fix the feedforward SNR to $-1$ dB and simulate each design multiple times. 
The BLER performances of various designs are presented in Fig.~\ref{fig:S1}. To showcase the impact of feature extractors, we do not incorporate belief network into GBAF code.

As can be seen from Fig.~\ref{fig:S1}, with the seven designs of feature extractor, GBAF code can achieve a BLER as low as $5\times 10^{-9}$ at a feedforward SNR of $-1$ dB. As far as the BLER performance is concerned, the most preferable architectures are designs C and D. On the other hand, as far as the computational resource is concerned, design C is twice more efficient than design D.
In the following, we will focus on designs C and D, and perform extensive simulations on a wide range of SNRs and compare their performances with other feedback codes.

\begin{figure}[t]
  \centering
  \includegraphics[width=0.8\columnwidth]{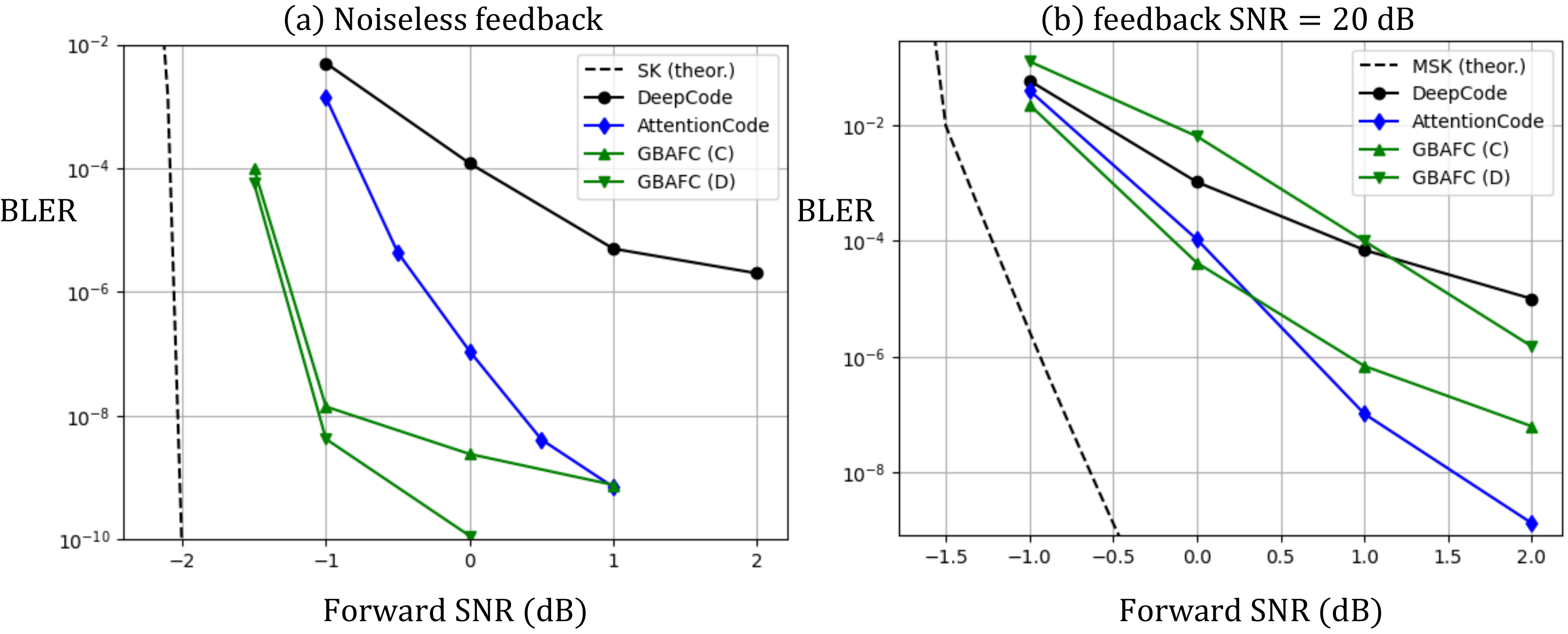}\\
  \caption{BLER versus feedforward SNR for various feedback codes. The system setup is $K=51$, $N = 153$, $m=3$, $\ell=17$.}
\label{fig:S2}
\end{figure}

Fig.~\ref{fig:S2} compares the BLER performance of different feedback codes with noiseless and noisy feedback.
In particular, we plot the theoretical performances of the SK and modulo-SK schemes \cite{SK1,ModuloSK} as benchmarks.
In the noiseless feedback case, GBAF code with either design C or design D achieves significant performance gains over AttentionCode. To attain a BLER of $10^{-8}$, GBAF code with designs C or D is about $1.2$ dB better than AttentionCode.
In the noisy feedback case (20 dB feedback SNR), the performance of GBAF code with design D deteriorates for a large margin. On the other hand, GBAF code with design C still performs well.

Based on the simulation results in this section, we choose design C as the default feature extractor for GBAF code. In the noiseless feedback case, design C is slightly worse than design D and outperforms all other designs. In the noisy feedback case, design C is much better than design D. In addition, design C is twice more efficient than design D as far as the required computational resource is concerned.

%%%%%%%%%%%%%%%%%%%%%%%%%%%%%%%%%%%%%%%%%
%%%%%%%%%%%%%%%%%%%%%%%%%%%%%%%%%%%%%%%%%
%%%%%%%%%%%%%%%%%%%%%%%%%%%%%%%%%%%%%%%%%
\subsection{ReLu versus GeLu}
In the machine learning community, GeLu activation function is widely considered as a better alternative to ReLu, thanks to its differentiability at $x=0$ for an input $x$. In our simulations, we find that the choice of the activation function is crucial to the performance of GBAF code, and GeLu is better than ReLu only when an ultra-low BLER is to be achieved. In this section, we perform extensive simulations to compare the performance of ReLu and GeLu when used in feature extractor (design C) of GBAF code.

\begin{figure}[t]
  \centering
  \includegraphics[width=0.8\columnwidth]{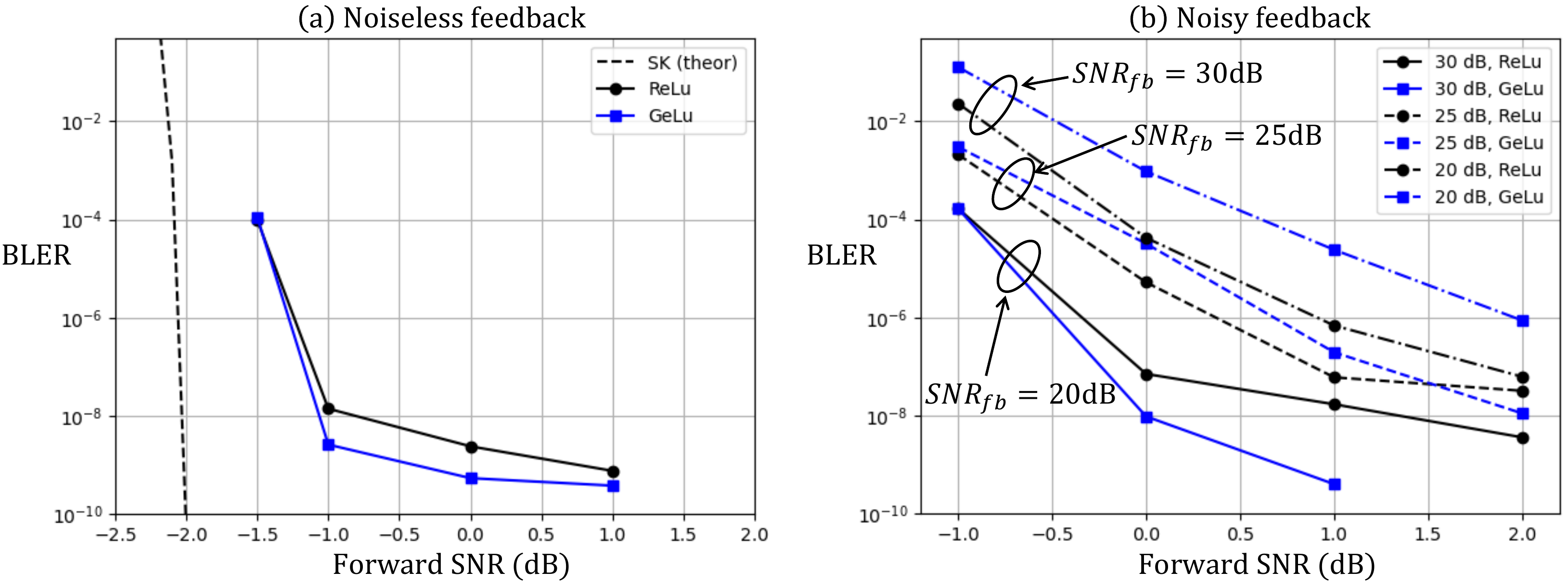}\\
  \caption{Performance comparison between ReLu and GeLu when used in feature extractor (design C) of GBAF code. The system setup is $K=51$, $N = 153$, $m=3$, $\ell=17$. (a) noiseless feedback, (b) noisy feedback.}
\label{fig:Sim_ReLuGeLu1}
\end{figure}

Fig.~\ref{fig:Sim_ReLuGeLu1}(a) compares the performance of ReLu and GeLu when used in the feature extractor, where the feedback is assumed to be noiseless. As can be seen, GeLu is beneficial to the BLER performance. Compared with ReLu, the performance of GBAF code is improved by $5$ times at a feedforward SNR of $-1$ dB when GeLu is used as the activation function.

Fig.~\ref{fig:Sim_ReLuGeLu1}(b) compares the performance of ReLu and GeLu in the noisy feedback setup, where we progressively decrease the feedback SNR from 30 dB to 20 dB.
As shown, when both the feedforward and feedback SNRs are large (e.g., $SNR_{ff}>1$ dB and $SNR_{fb}\geq 25$ dB), GeLu exhibits better performance than ReLu.
In contrast, when the feedforward and feedback SNRs decrease, ReLu performs better.
As a conclusion, if an ultra-low BLER (e.g., BLER lower than $10^{-7}$) is to be achieved, GeLu is a better choice than ReLu.

\begin{figure}[t]
  \centering
  \includegraphics[width=0.8\columnwidth]{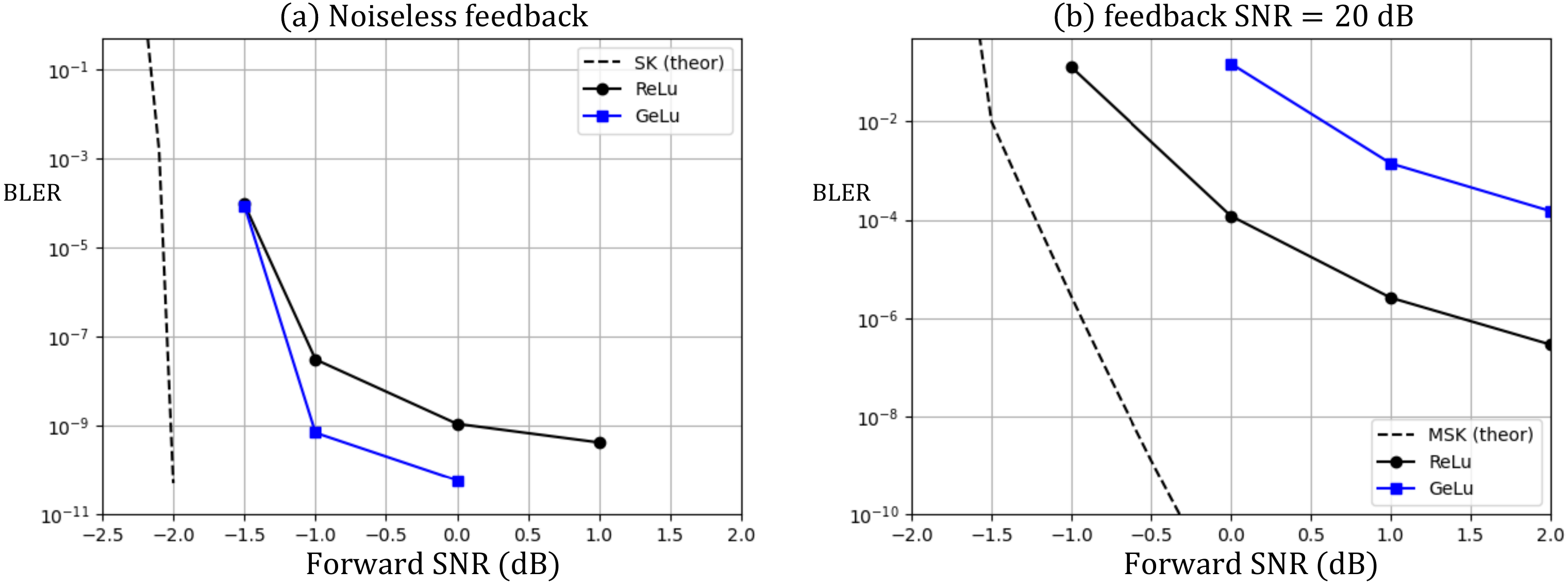}\\
  \caption{Performance comparison between ReLu and GeLu when used in feature extractor (design C) of GBAF code. The system setup is $K=51$, $N = 153$, $m=3$, $\ell=17$. GBAF code uses both feature extractor and belief network. (a) noiseless feedback, (b) $20$ dB noisy feedback.}
\label{fig:Sim_ReLuGeLu3}
\end{figure}

To confirm the above results, we perform additional simulations by taking the belief network into account. Specifically, we consider GBAF code with both feature extractor (design C) and belief network. When ReLu and GeLu are used as the activation function of the feature extractor, Fig.~\ref{fig:Sim_ReLuGeLu3} compares the performance achieved by GBAF code.
When the feedback channel is noiseless, GeLu yields much lower BLER than ReLu.
When the feedback channel is noisy ($20$ dB), on the other hand, ReLu is better choice than GeLu.
This confirms our conclusions drawn from Fig.~\ref{fig:Sim_ReLuGeLu1}.

To conclude this appendix, the default feature extractor for GBAF code is set to be design C with the ReLu activation function. An alternative to ReLu is GeLu when the target BLER is ultra low (e.g., lower than $10^{-7}$).
% In AWGN channels, the best performance of GBAFC with noiseless feedback is shown in Fig. \ref{fig:Sim_ReLuGeLu3}(a); the best performance of GBAFC with $20$ dB noisy feedback is shown in Fig. \ref{fig:S2}(b).

\bibliographystyle{IEEEtran}
\bibliography{IEEEabrv,ref.bib}

\ifCLASSOPTIONcaptionsoff
  \newpage
\fi



%\vfill


\end{document}